\documentclass[11pt]{article}
\usepackage{latexsym,amssymb,amsmath}
\usepackage[dvips]{graphics,graphicx,epsfig,color}
\begin{document}
\thispagestyle{empty}
\title{\bf  Three methods for the description of the temporal response to a SH plane
impulsive seismic wave in a soft elastic layer overlying a hard
elastic substratum}
\author{\bf Armand Wirgin \thanks {Laboratoire de M\'ecanique et d'Acoustique, UPR 7051 du CNRS,
 31 chemin Joseph Aiguier, 13009 Marseille, France.}}
\date{\today}
\maketitle
\begin{abstract}
We treat the case of a flat stress-free surface (i.e., the ground
in seismological applications) separating air from a homogeneous,
isotropic, solid substratum overlain by a homogeneous, isotropic,
solid layer (in contact with the ground)  solicited by a SH plane
body wave incident in the substratum. The analysis is first
carried out in the frequency domain and subsequently in the time
domain. The frequency domain response is {\it normal} in that no
resonances are excited (a resonance is here understood to be a
situation in which the response is infinite in the absence of
dissipation). The translation of this in the time domain is that
the scattered pulse is of relatively-short duration. The duration
of the pulse is shown to be largely governed by radiation damping
which shows up in the imaginary parts of the complex
eigenfrequencies of the configuration. Three methods are
elaborated for the computation of the time history and give rise
to the same numerical solutions for a large variety of
configurations of interest in the geophysical setting under the
hypothesis of non-dissipative, dispersionless media. The method
appealing to the complex eigenfrequency representation is shown to
be the simplest and most physically-explicit way of obtaining the
time history (under the same hypothesis). Moreover, it is
particularly suited for the case in which modes can be excited as
occurs when the incident wave is not plane or the boundary
condition is not of the stress-free variety for all transverse
coordinates on the ground plane.
\end{abstract}
\newpage
\tableofcontents
\newpage
\newpage
\section{Introduction}
This work is inspired by the problem of predicting the effects of
earthquakes in cities. It is known that the most dangerous effects
are produced in cities built on soft underground underlain by a
hard substratum. A simple model of the city is considered herein
in which the buildings are absent (i.e., the ground is flat), the
soft underground is constituted by a homogeneous, soft layer
overlying, and in welded contact with, a homogeneous, hard
substratum. This configuration is solicited by a SH plane body
wave and the object is to determine the time history of response
on the ground, preferably in a numerically-efficient,
physically-understandable manner.
\section{Space-time and space-frequency formulations}
In the following, we shall be concerned with the determination of
the vectorial displacement field $\mathbf{u}$ on, and underneath,
the ground in response to a seismic solicitation. In general,
$\mathbf{u}$ is a function of the spatial coordinates, incarnated
in the vector $\mathbf{x}$ and time $t$, so that
$\mathbf{u}=\mathbf{u}(\mathbf{x},t)$.

We first carry out our analysis in the frequency domain, and thus
search for $\mathbf{u}(\mathbf{x},\omega)$, with $\omega$ the
angular frequency.

Fourier analysis tells us that $\mathbf{u}(\mathbf{x},t)$ and
$\mathbf{u}(\mathbf{x},\omega)$ are related by
\begin{equation}\label{fourier}
  \mathbf{u}(\mathbf{x},t)=\int_{-\infty}^{\infty}\mathbf{u}(\mathbf{x},\omega)\exp(-i\omega
  t)d\omega~,
\end{equation}
wherein it should be noted that $\mathbf{u}(\mathbf{x},\omega)$ is
a generally-complex function, whereas $\mathbf{u}(\mathbf{x},t)$
is a real function. The second step will therefore deal with the
computation of the integral in (\ref{fourier}).
\section{Frequency domain analysis of the reflection of a SH plane body wave
from a stress-free planar
boundary overlying a soft layer underlain by a hard
substratum}\label{sec54}
%
\subsection{Features of the problem}
\begin{figure}
[ptb]
\begin{center}
\includegraphics[scale=0.5] {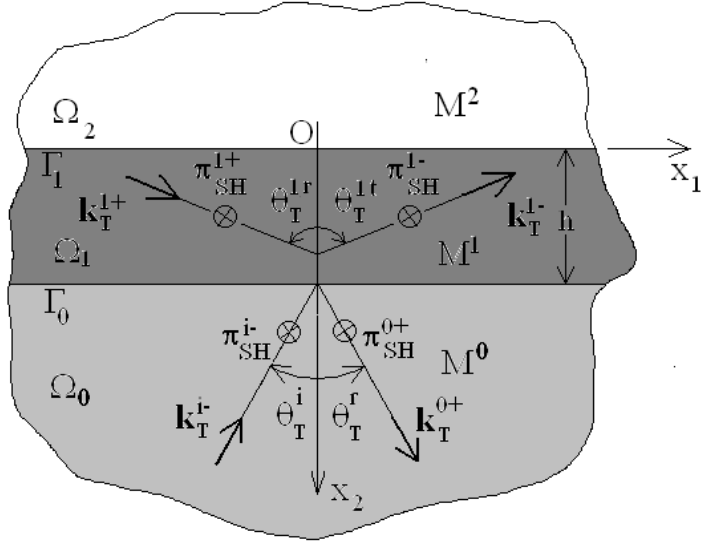}
  \caption{Cross section view of the configuration of  a stress-free flat surface overlying a soft layer
  underlain by a hard solid substratum submitted to a SH plane
  wave, propagating initially in the substratum.}
  \label{fig5}
  \end{center}
\end{figure}

Since everything is invariant with $x_{3}$, the analysis  takes
place in the $x_{1}-x_{2}$ (sagittal) plane depicted in fig.
\ref{fig5}.

In this figure: $\Gamma_{0}$ designates the (trace of the)
interface between the substratum (half-space domain $\Omega_{0}$)
and the soft layer (laterally-unbounded domain $\Omega_{1}$), and
$\Gamma_{1}$ designates the (trace of the) flat ground. The medium
$M^{2}$ above the latter is air, assumed to be the vacumn for the
purpose of the analysis. The media in the layer and substratum are
the elastic (or viscoelastic) solids $M^{1}$ and  $M^{0}$
respectively.

The incident plane wave propagates in $\Omega_{0}$ toward the
 the interface $\Gamma_{0}$ and the ground  $\Gamma_{1}$.
Since the latter is stress-free (i.e., the normal and tangential
components of traction are nil on the boundary), the total
displacement field vanishes in the region $\Omega_{2}$ above the
boundary (see fig. \ref{fig5}).
One can always choose the cartesian coordinate system so that the
wavevector associated with the incident shear wave (the subscript
$T$ will constitute a reminder that we deal with shear=transverse
waves in the following) lies in the $x_{1}-x_{2}$ plane. This is
assumed herein and signifies that the displacement associated with
this wave is perpendicular to the $x_{1}-x_{2}$ plane and
therefore lies in a horizontal plane. Thus, the incident wave is a
shear wave and the associated displacement is horizontal; i.e., a
shear-horizontal (SH) wave. Moreover, the motion associated with
this wave is, due to the choice of the cartesian reference system,
independent of the coordinate $x_{3}$. This implies that the
resultant total motion induced by this incident wave is
independent of $x_{3}$, i.e., the boundary value problem is 2D, so
that it is sufficient to look for the displacement field in the
$x_{1}-x_{2}$ plane. Actually, since we already know that the
total displacement vanishes in the half plane above the boundary
we must look for the total displacement field (hereafter
designated by $\mathbf{u}^{0}(\mathbf{x},\omega)$) only in
 $\Omega_{0}$ and  $\Omega_{1}$.

Hereafter, we designate the density and Lam\'e parameters in
$\Omega_{j}$ by $\rho^{j}$ and $\lambda^{j},~\mu^{j}$ (for
$j=0,1$) respectively.
%
\subsection{Governing equations}
The mathematical translation of the boundary value problem in the
{\it space-frequency domain} is:
\begin{multline}\label{sevenhunthirteen}
  \mu u^{l}_{j,\beta\beta}(\mathbf{x},\omega)+(\lambda^{l}+
  \mu^{l})u^{l}_{\beta,\beta j}(\mathbf{x},\omega)+
  \rho^{l}\omega^{2}
  u^{l}_{j}(\mathbf{x},\omega)=0~;~\forall\mathbf{x}=(x_{1},x_{2})\in\Omega_{l}
  ~~;
  \\
  ~~j=1,2,3~~,~~l=0,1~~,~~\beta=1,2~,
\end{multline}
\begin{equation}\label{sevenhunthirteenab}
 u^{l}_{j,3}(\mathbf{x},\omega)=0~;~\forall\mathbf{x}\in\Omega_{l}~~;~~j=1,2,3~~,~~l=0,1~,
\end{equation}
\begin{equation}\label{sevenhunfourteen}
(\lambda^{1}+2\mu^{1})u_{2,2}^{1}(\mathbf{x},\omega)+\lambda^{1}u_{1,1}^{1}(\mathbf{x},\omega)=0
~~\text{on}~\Gamma_{1}~,
\end{equation}
\begin{equation}\label{sevenhunfourteena}
\mu^{1}u_{3,2}^{1}(\mathbf{x},\omega)= ~~\text{on}~\Gamma_{1}~,
\end{equation}
\begin{equation}\label{sevenhunfourteenb}
-\mu^{1}(u_{1,2}^{1}(\mathbf{x},\omega)+u_{2,1}^{1}(\mathbf{x},\omega))=0
~~\text{on}~\Gamma_{1}~.
\end{equation}
\begin{equation}\label{sevenhunthirtyeight25}
u_{2}^{0}(\mathbf{x},\omega)=u_{2}^{1}(\mathbf{x},\omega)~~\text{on}~\Gamma_{0}~,
\end{equation}
\begin{equation}\label{sevenhunthirtyeight26}
u_{3}^{0}(\mathbf{x},\omega)=u_{3}^{1}(\mathbf{x},\omega)~~\text{on}~\Gamma_{0}~,
\end{equation}
\begin{equation}\label{sevenhunthirtyeight27}
 -u_{1}^{0}(\mathbf{x},\omega)=-u_{1}^{1}(\mathbf{x},\omega)~~\text{on}~\Gamma_{0}~,
\end{equation}
\begin{equation}\label{sevenhunthirtyeight28}
(\lambda^{0}+2\mu^{0})u_{2,2}^{0}(\mathbf{x},\omega)+\lambda^{0}u_{1,1}^{0}(\mathbf{x},\omega)=
(\lambda^{1}+2\mu^{1})u_{2,2}^{1}(\mathbf{x},\omega)+\lambda^{1}u_{1,1}^{1}(\mathbf{x},\omega)
~~\text{on}~\Gamma_{0}~,
\end{equation}
\begin{equation}\label{sevenhunthirtyeight29}
\mu^{0}u_{3,2}^{0}(\mathbf{x},\omega)=
\mu^{1}u_{3,2}^{0}(\mathbf{x},\omega)~~\text{on}~\Gamma_{0}~,
\end{equation}
\begin{equation}\label{sevenhunthirtyeight30}
-\mu^{0}(u_{1,2}^{0}(\mathbf{x},\omega)+u_{2,1}^{0}(\mathbf{x},\omega))=
-\mu^{1}(u_{1,2}^{1}(\mathbf{x},\omega)+u_{2,1}^{1}(\mathbf{x},\omega))~~\text{on}~\Gamma_{0}~.
\end{equation}
\begin{multline}\label{sevenhunfourtyone}
u_{j}^{ld}(\mathbf{x},\omega):=
u_{j}^{l}(\mathbf{x},\omega)-\delta_{l0}u_{0}^{i}(\mathbf{x},\omega)\sim
\text{outgoing waves}~;
\\
~\|\mathbf{x}\|\rightarrow\infty~,~~\mathbf{x}\in\Omega_{j}~,~j=0,1~,~l=0,1~,
\end{multline}
wherein $u_{j}^{ld}(\mathbf{x},\omega)$ is the (unknown)
diffracted field in $\Omega_{l}$, $\delta_{jk}$ the Kronecker
delta symbol, and:
\begin{equation}\label{sevenhunfourtytwo}
u^{i}_{3}(\mathbf{x},\omega)=A^{i-}_{3}\exp(i\mathbf{k}^{i-}_{T}\cdot\mathbf{x})~,~~
u^{i}_{1}(\mathbf{x},\omega)~=~u^{i}_{2}(\mathbf{x},\omega)=0~;
~\forall\mathbf{x}\in\Omega_{0}~,
\end{equation}
\begin{equation}\label{sevenhunfourtythree}
\mathbf{k}^{i-}_{T}=(k^{i}_{T1},~k^{i-}_{T2})~,~k^{i}_{T1}=
k^{0}_{T}\sin\theta_{T}^{i}~,~k^{i-}_{T2}=-k^{0}_{T}\cos\theta_{T}^{i}~,~
k^{0}_T=\frac{\omega}{\sqrt{\frac{\mu^{0}}{\rho^{0}}}}~,
\end{equation}
$\theta_{T}^{i}$ being the angle of incidence with respect to the
$x_{2}$ axis.

Eq. (\ref{sevenhunthirteen}) is the space-frequency domain
equation(s) of motion,
(\ref{sevenhunfourteen})-(\ref{sevenhunthirtyeight30}) the
boundary condition(s), (\ref{sevenhunfourtyone}) the radiation
condition, and
(\ref{sevenhunfourtytwo})-(\ref{sevenhunfourtythree}) the
description of the incident wave.

Until further notice, we drop the $\omega-$dependence on all field
quantities and consider it to be implicit.
\subsubsection{Field representations incorporating the radiation condition}
As in the previous section, and on account of the outgoing wave
condition(s) (\ref{sevenhunfourtyone}), we adopt the following
field representations:
\begin{equation}\label{sevenhunfourtyfour}
u_{1}^{0d}=i\int_{-\infty}^{\infty}
[\Phi^{0+}(k_{1})k_{1}\exp(i\mathbf{k}^{0+}_{L} \cdot\mathbf{x})+
\Psi_{3}^{0+}(k_{1})k_{T2}^{0+}\exp(i\mathbf{k}^{0+}_{T}
\cdot\mathbf{x})]dk_{1}~,
\end{equation}
\begin{equation}\label{sevenhunfourtyfive}
u_{2}^{0d}=i\int_{-\infty}^{\infty}
[\Phi^{0+}(k_{1})k_{L2}^{0+}\exp(i\mathbf{k}^{0+}_{L}
\cdot\mathbf{x})-
\Psi_{3}^{0+}(k_{1})k_{1}\exp(i\mathbf{k}^{0+}_{T}
\cdot\mathbf{x})]dk_{1}~,
\end{equation}
\begin{equation}\label{sevenhunfourtysix}
u_{3}^{0d}=\int_{-\infty}^{\infty}
\Xi_{3}^{0+}(k_{1})\exp(i\mathbf{k}^{0+}_{T}
\cdot\mathbf{x})dk_{1}~,
\end{equation}
\begin{multline}\label{sevenhunfourtyseven}
u_{1}^{1}=u_{1}^{1d}=i\int_{-\infty}^{\infty}
\big[\Phi^{1-}(k_{1})k_{1}\exp(i\mathbf{k}^{1-}_{L}
\cdot\mathbf{x})+
\Psi_{3}^{1-}(k_{1})k_{T2}^{1-}\exp(i\mathbf{k}^{1-}_{T}
\cdot\mathbf{x})+
\\
\Phi^{1+}(k_{1})k_{1}\exp(i\mathbf{k}^{1+}_{L} \cdot\mathbf{x})+
\Psi_{3}^{1+}(k_{1})k_{T2}^{1+}\exp(i\mathbf{k}^{1+}_{T}
\cdot\mathbf{x})\big]dk_{1}~,
\end{multline}
\begin{multline}\label{sevenhunfourtyeight}
u_{2}^{1}=u_{2}^{1d}=i\int_{-\infty}^{\infty}
\big[\Phi^{1-}(k_{1})k_{L2}^{1-}\exp(i\mathbf{k}^{1-}_{L}
\cdot\mathbf{x})-
\Psi_{3}^{1-}(k_{1})k_{1}\exp(i\mathbf{k}^{1-}_{T}
\cdot\mathbf{x})+
\\
\Phi^{1+}(k_{1})k_{L2}^{1+}\exp(i\mathbf{k}^{1+}_{L}
\cdot\mathbf{x})-
\Psi_{3}^{1+}(k_{1})k_{1}\exp(i\mathbf{k}^{1+}_{T}
\cdot\mathbf{x})\big]dk_{1}~,
\end{multline}
\begin{equation}\label{sevenhunfourtynine}
u_{3}^{1}=u_{3}^{1d}=\int_{-\infty}^{\infty}\big[
\Xi_{3}^{1-}(k_{1})\exp(i\mathbf{k}^{1-}_{T}
\cdot\mathbf{x})+\Xi_{3}^{1+}(k_{1})\exp(i\mathbf{k}^{1+}_{T}
\cdot\mathbf{x})\big] dk_{1}~,
\end{equation}
wherein
\begin{equation}\label{sevenhunfifty}
\mathbf{k}^{j\pm}_{L}=(k_{1},k^{j\pm}_{L2})~~,~~k^{j\pm}_{L2}=\pm\sqrt{\left(k_{L}^{j}\right)
^{2}-\left( k_{1}\right) ^{2}}~~,~~\Re\sqrt{\left(k_{L}^{j}\right)
^{2}-\left( k_{1}\right) ^{2}}\geq
0~,~\Im\sqrt{\left(k_{L}^{j}\right) ^{2}-\left( k_{1}\right)
^{2}}\geq 0 ~,
\end{equation}
\begin{equation}\label{sevenhunfiftyone}
\mathbf{k}^{j\pm}_{T}=(k_{1},k^{j\pm}_{T2})~~,~~k^{j\pm}_{T2}=\pm\sqrt{\left(k_{T}^{j}\right)
^{2}-\left( k_{1}\right) ^{2}}~~,~~\Re\sqrt{\left(k_{T}^{j}\right)
^{2}-\left( k_{1}\right) ^{2}}\geq
0~,~\Im\sqrt{\left(k_{T}^{j}\right) ^{2}-\left( k_{1}\right)
^{2}}\geq 0 ~,
\end{equation}
with
\begin{equation}\label{sevenhunfiftytwo}
k^{j}_{L}=\frac{\omega}{c^{j}_{L}}=
\frac{\omega}{\sqrt{\frac{\lambda^{j}+2\mu^{j}}{\rho^{j}}}}~~,~~
k^{j}_{T}=
\frac{\omega}{c^{j}_{T}}=\frac{\omega}{\sqrt{\frac{\mu^{j}}{\rho^{j}}}}~.
\end{equation}
Note that these field representations involve nine unknown
functions $\Phi^{0+},~\Psi^{0+},~\Xi^{0+}$, $
\Phi^{1-},~\Psi^{1-},~\Xi^{1-}$, $\Phi^{1+},~\Psi^{1+},~\Xi^{1+}
$. The latter will be obtained by applying the nine boundary
conditions embodied in
(\ref{sevenhunfourteen})-(\ref{sevenhunthirtyeight30}).
\subsection{Application of the boundary condition(s)}
The use of (\ref{sevenhunfourteen}), (\ref{sevenhunfourteenb}),
(\ref{sevenhunthirtyeight25}), (\ref{sevenhunthirtyeight27}),
(\ref{sevenhunthirtyeight28}), and (\ref{sevenhunthirtyeight30}),
 in
(\ref{sevenhunfourtyfour})-(\ref{sevenhunfiftytwo}) gives rise to:
\begin{equation}\label{twohunthirtytwo}
\Phi^{0+}=\Psi_{3}^{0+}=\Phi^{1-}=\Psi_{3}^{1-}=\Phi^{1+}=\Psi_{3}^{1+}=0~~;~~\forall
k_{1}\in\mathbb{R}~.
\end{equation}
The next step consists in using (\ref{sevenhunfourteena}) in
 (\ref{sevenhunfourtynine}) to
obtain:
\begin{equation}\label{sevenhunfiftythree}
i\mu^{1}\int_{-\infty}^{\infty}\big[ k_{2}^{1-}\Xi_{3}^{1-}(k_{1})
+k_{2}^{1+}\Xi_{3}^{1+}(k_{1})\big]
\exp(ik_{1}x_{1})dk_{1}=0~~;\forall x_{1}\in\mathbb{R}~.
\end{equation}
By Fourier inversion we get
\begin{equation}\label{sevenhunfiftyfour}
\Xi_{3}^{1-}=\Xi_{3}^{1+}:=A^{1}~~;\forall k_{1}\in\mathbb{R}~,
\end{equation}
whence
\begin{equation}\label{sevenhunfiftyfive}
u^{1}(\mathbf{x})=2\int_{-\infty}^{\infty}A^{1}(k_{1})\cos(k_{2}^{1}x_{2})
\exp(ik_{1}x_{1})dk_{1}=~~;\forall \mathbf{x}\in\Omega_{1}~,
\end{equation}
wherein
\begin{equation}\label{sevenhunfiftysix}
k_{2}^{j}:=k_{T2}^{j+}~~;~~j=0,1~~,~~k_{2}^{i}:=k_{T2}^{i+}=-k_{2}^{i-}~.
\end{equation}
By the same token, we can write (\ref{sevenhunfourtynine}) as
\begin{equation}\label{sevenhunfiftyseven}
u^{0d}(\mathbf{x})=\int_{-\infty}^{\infty}A^{0+}(k_{1})\exp(ik_{2}^{0+}x_{2})
\exp(ik_{1}x_{1})dk_{1}=~~;\forall \mathbf{x}\in\Omega_{0}~,
\end{equation}
so that using (\ref{sevenhunfiftyfive}) and
(\ref{sevenhunfiftyseven}) together with
(\ref{sevenhunthirtyeight26}) and (\ref{sevenhunthirtyeight29})
leads to:
\begin{multline}\label{sevenhunfiftysevena}
A^{i-}\exp(-ik_{2}^{i}h)\exp(ik^{i}x_{1})+\int_{-\infty}^{\infty}
A^{0+}(k_{1})\exp(ik_{2}^{0}h)\exp(ik_{1}x_{1})dk_{1}-
\\
\int_{-\infty}^{\infty}
2A^{1}(k_{1})\cos(k_{2}^{1}h)\exp(ik_{1}x_{1})dk_{1}=0~~;\forall
x_{1}\in\mathbb{R}~,
\end{multline}
\begin{multline}\label{sevenhunfiftyeight}
-i\mu^{0}k_{2}^{i}A^{i-}\exp(-ik_{2}^{i}h)\exp(ik^{i}x_{1})+\int_{-\infty}^{\infty}
i\mu^{0}k_{2}^{0}A^{0+}(k_{1})\exp(ik_{2}^{0}h)\exp(ik_{1}x_{1})dk_{1}+
\\
\int_{-\infty}^{\infty}
\mu^{1}k_{2}^{1}2A^{1}(k_{1})\sin(k_{2}^{1}h)\exp(ik_{1}x_{1})dk_{1}=0~~;\forall
x_{1}\in\mathbb{R}~.
\end{multline}
By Fourier inversion we find:
\begin{equation}\label{sevenhunfiftynine}
A^{i-}\exp(-ik_{2}^{i}h)\delta(k_{1}-k_{1}^{i})+
A^{0+}(k_{1})\exp(ik_{2}^{0}h)-
2A^{1}(k_{1})\cos(k_{2}^{1}h)=0~~;\forall k_{1}\in\mathbb{R}~,
\end{equation}
\begin{equation}\label{sevenhunsixty}
-i\mu^{0}k_{2}^{i}A^{i-}\exp(-ik_{2}^{i}h)\delta(k_{1}-k_{1}^{i})+
i\mu^{0}k_{2}^{0}A^{0+}(k_{1})\exp(ik_{2}^{0}h)+
\mu^{1}k_{2}^{1}2A^{1}(k_{1})\sin(k_{2}^{1}h)=0~~;\forall
x_{1}\in\mathbb{R}~.
\end{equation}
This system of two equations can be written as the matrix equation
\begin{equation}\label{sevenhunsixtyone}
\begin{pmatrix}
\exp(ik_{2}^{0}h) & -\cos(k_{2}^{1}h)
\\
i\mu^{0}k_{2}^{0}\exp(ik_{2}^{0}h) &
\mu^{1}k_{2}^{1}\sin(k_{2}^{1}h)
\end{pmatrix}
\begin{pmatrix}
A^{0+}
\\
2A^{1}
\end{pmatrix}
=
\begin{pmatrix}
-A^{i-}\exp(-ik_{2}^{i}h)\delta(k_{1}-k_{1}^{i})
\\
i\mu^{0}k_{2}^{i}A^{i-}\exp(-ik_{2}^{i}h)\delta(k_{1}-k_{1}^{i})
\end{pmatrix}
~.
\end{equation}
It follows that:
\begin{equation}\label{sevenhunsixtytwo}
  A^{0+}(k_{1})=\mathcal{R}^{0}(k_{1})A^{i-}\delta(k_{1}-k_{1}^{i})~,
\end{equation}
wherein
\begin{equation}\label{sevenhunsixtythree}
  \mathcal{R}^{0}(k_{1})=\left[ \frac{\mu^{0}k_{2}^{i}\cos(k_{2}^{1}h)+i\mu^{1}k_{2}^{1}\sin(k_{2}^{1}h)}
  {\mu^{0}k_{2}^{i}\cos(k_{2}^{1}h)-i\mu^{1}k_{2}^{1}\sin(k_{2}^{1}h)}\right] \exp[-i(k_{2}^{i}+k_{2}^{0})h]~,
\end{equation}
and
\begin{equation}\label{sevenhunsixtyfour}
  A^{1}(k_{1})=\frac{1}{2}\mathcal{R}^{1}(k_{1})A^{i-}\delta(k_{1}-k_{1}^{i})~,
\end{equation}
wherein
\begin{equation}\label{sevenhunsixtyfive}
  \mathcal{R}^{1}(k_{1})=\left[ \frac{\mu^{0}k_{2}^{0}+\mu^{0}k_{2}^{i}}
  {\mu^{0}k_{2}^{i}\cos(k_{2}^{1}h)-i\mu^{1}k_{2}^{1}\sin(k_{2}^{1}h)}\right] \exp(-ik_{2}^{i}h)
  ~.
\end{equation}
\subsection{The scattered field}
The consequence of all this is that:
\begin{equation}\label{sevenhunsixtysix}
  u_{1}^{0d}(\mathbf{x},\omega)=u_{2}^{0d}(\mathbf{x},\omega)=0~~
  ;~~\forall \mathbf{x} \in\Omega_{0}~,
\end{equation}
\begin{equation}\label{sevenhunsixtysixa}
 u_{3}^{0d}(\mathbf{x},\omega)=\int_{-\infty}^{\infty}A^{0+}
  e^{i(k_{1}x_{1}+k_{2}^{0}x_{2})}dk_{1}=\mathcal{R}^{0}(k_{1}^{i})A^{i-} e^{i(k_{1}^{i}x_{1}+k_{2}^{0i}x_{2})}~~
  ;~~\forall \mathbf{x} \in\Omega_{0}~,
\end{equation}
\begin{equation}\label{sevenhunsixtyseven}
   u_{1}^{1}(\mathbf{x},\omega)=u_{1}^{1d}(\mathbf{x},\omega)=
   u_{2}^{1d}(\mathbf{x},\omega)= u_{2}^{1}(\mathbf{x},\omega)=0~~,
  ;~~\forall \mathbf{x} \in\Omega_{1}~,
\end{equation}
\begin{equation}\label{sevenhunsixtysevena}
u_{3}^{1}(\mathbf{x},\omega)=u_{3}^{1d}(\mathbf{x},\omega)=2\int_{-\infty}^{\infty}A^{1}
  e^{ik_{1}x_{1}}\cos(k_{2}^{0}x_{2})dk_{1}=
  \mathcal{R}^{1}(k_{1}^{i})A^{i-}e^{ik_{1}^{i}x_{1}}\cos(k_{2}^{1i}x_{2})
  ~~
  ;~~\forall \mathbf{x} \in\Omega_{1}~,
\end{equation}
wherein
\begin{equation}\label{sevenhunsixtysevenb}
k_{2}^{ji}:=\sqrt{(k^{j})^{2}-(k_{1}^{i})^{2}}=\sqrt{(k^{j})^{2}-(k^{0}\sin\theta_{T}^{i})^{2}}~~,~~\Re
k_{2}^{ji}\geq 0~,~~\Im k_{2}^{ji}\geq
0~~,~~j=0,1~~(k_{2}^{0i}=k_{2}^{i})~,
\end{equation}
\begin{equation}\label{sevenhunsixtysevenc}
  \mathcal{R}^{0}(k_{1}^{i})=\left[ \frac{\mu^{0}k_{2}^{0i}\cos(k_{2}^{1i}h)+i\mu^{1}k_{2}^{1}\sin(k_{2}^{1i}h)}
  {\mu^{0}k_{2}^{0i}\cos(k_{2}^{1i}h)-i\mu^{1}k_{2}^{1i}\sin(k_{2}^{1i}h)}\right] \exp[-2ik_{2}^{0i}h]~,
\end{equation}
and
\begin{equation}\label{sevenhunsixtysevend}
  \mathcal{R}^{1}(k_{1}^{i})=\left[ \frac{2\mu^{0}k_{2}^{0i}}
  {\mu^{0}k_{2}^{0i}\cos(k_{2}^{1i}h)-i\mu^{1}k_{2}^{1i}\sin(k_{2}^{1i}h)}\right]
  \exp[-ik_{2}^{0i}h]
  ~.
\end{equation}
 These results, together with (\ref{sevenhunfourtytwo}),
indicate that the diffracted fields in $\Omega_{0}$ and
$\Omega_{1}$ have the same (SH) polarization ($\boldsymbol{\pi}$
in fig. \ref{fig5}) as the incident field.
\subsection{Total fields in the two media}
The preceding results show that the fields in the two media can be
decomposed as follows:
\begin{equation}\label{sevenhunsixtyeight}
  u_{3}^{0}(\mathbf{x},\omega)=u_{3}^{0-}(\mathbf{x},\omega)+u_{3}^{0+}(\mathbf{x},\omega)
  ~~
  ;\forall \mathbf{x} \in\Omega_{0}~,
\end{equation}
wherein
\begin{equation}\label{sevenhunsixtyeighta}
 u_{3}^{0-}(\mathbf{x},\omega)=u_{3}^{i}(\mathbf{x},\omega)=
 A^{i-}e^{i[k_{1}^{i}x_{1}-k_{2}^{0i}x_{2}]}
 ~,
\end{equation}
\begin{equation}\label{sevenhunsixtyeightb}
  u_{3}^{0+}(\mathbf{x},\omega)=
  \mathcal{R}^{0}(k_{1}^{i})A^{i-}e^{i[k_{1}^{i}x_{1}+k_{2}^{0i}x_{2}]}~,
\end{equation}
 and
\begin{equation}\label{sevenhunsixtynine}
  u_{3}^{1}(\mathbf{x},\omega)=
  u_{3}^{1-}(\mathbf{x},\omega)+u_{3}^{1+}(\mathbf{x},\omega)
  ~~
  ;\forall \mathbf{x} \in\Omega_{1}~
\end{equation}
wherein
\begin{equation}\label{sevenhunseventy}
 u_{3}^{1-}(\mathbf{x},\omega)
 :=\frac{\mathcal{R}^{1}(k_{1}^{i})}{2}A^{i-}e^{i[k_{1}^{i}x_{1}-k_{2}^{1i}x_{2}]}~,
\end{equation}
\begin{equation}\label{sevenhunseventyone}
 u_{3}^{1+}(\mathbf{x},\omega)
:
=\frac{\mathcal{R}^{1}(k_{1}^{i})}{2}A^{i-}e^{i[k_{1}^{i}x_{1}+k_{2}^{1i}x_{2}]}~.
\end{equation}
\newline
{\it Remark}
\newline
These results indicate that the diffracted field in $\Omega_{0}$
reduces to a specularly-reflected  wave $u_{3}^{0+}$ and the
diffracted (as well as total) field in $\Omega_{1}$ reduces to a
sum of a refracted-reflected  wave  $u_{3}^{1+}$ and a
refracted-transmitted wave  $u_{3}^{1-}$.
\newline
\newline
{\it Remark}
\newline
$u_{3}^{0-}$ and $u_{3}^{0+}$ are plane homogeneous (body) waves
so that the field in the half-space underneath the layer is
composed of two body waves.
\newline
\newline
{\it Remark}
\newline
  Since we
consider only the case of geological interest in which the
substratum is harder than the layer, the S-wave phase velocity in
the substratum is larger than the S-wave phase velocity in the
layer, which means that $k^{0}<k^{1}$ and consequently
$k_{2}^{1i}$ is real, so that both $u_{3}^{1-}$ and $u_{3}^{1+}$
are body waves. Thus, the field in the layer is also composed of
two body waves.
\newline
\newline
{\it Remark}
\newline
An important corollary of the previous remarks is (in the case of
geological interest) that an incident (plane) body wave in the
substratum can only excite (plane) body waves in both the
substratum and the layer.  Thus, if we want a surface wave to be
excited somewhere underneath the ground, then we have to introduce
some sort of modification of either the excitation, the nature of
the media, or the nature of the boundary condition on the ground.
\subsection{Numerical results for the frequency domain response in the layer}
\begin{figure}
[ptb]
\begin{center}
\includegraphics[scale=0.5] {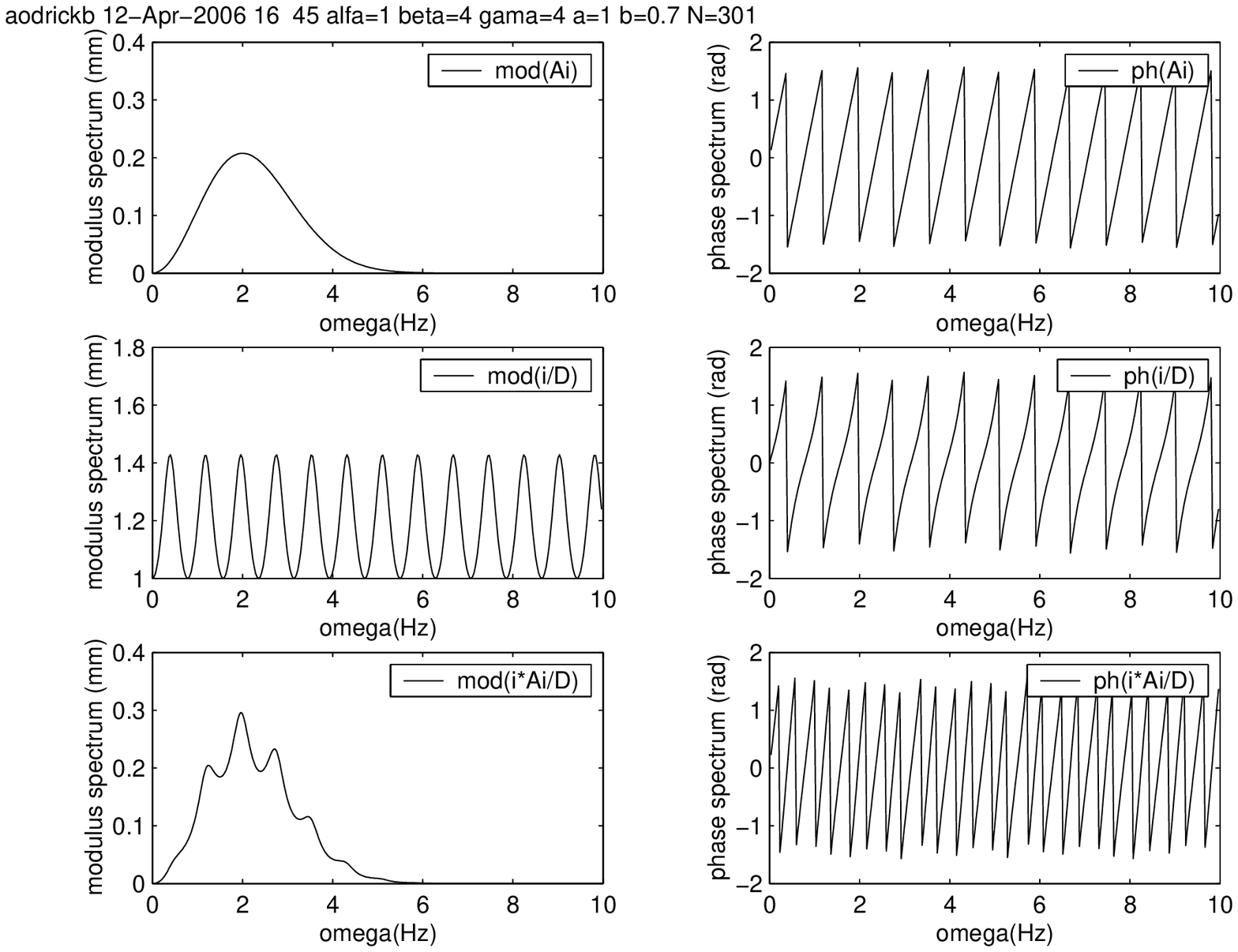}
  \caption{Spectrum of displacement response at $\mathbf{x}=(0,0)$.
  $\mathcal{A}=1$,
  $\alpha=1$, $\beta=4$, $\gamma=4$, $a=1$, $b=0.7$,
  corresponding to a case of separated pulses. The left-hand
  curves
  pertain to moduli, and the right hand curves to phases of the spectra.}
  \label{aodrickb1}
  \end{center}
\end{figure}

\begin{figure}
[ptb]
\begin{center}
\includegraphics[scale=0.5] {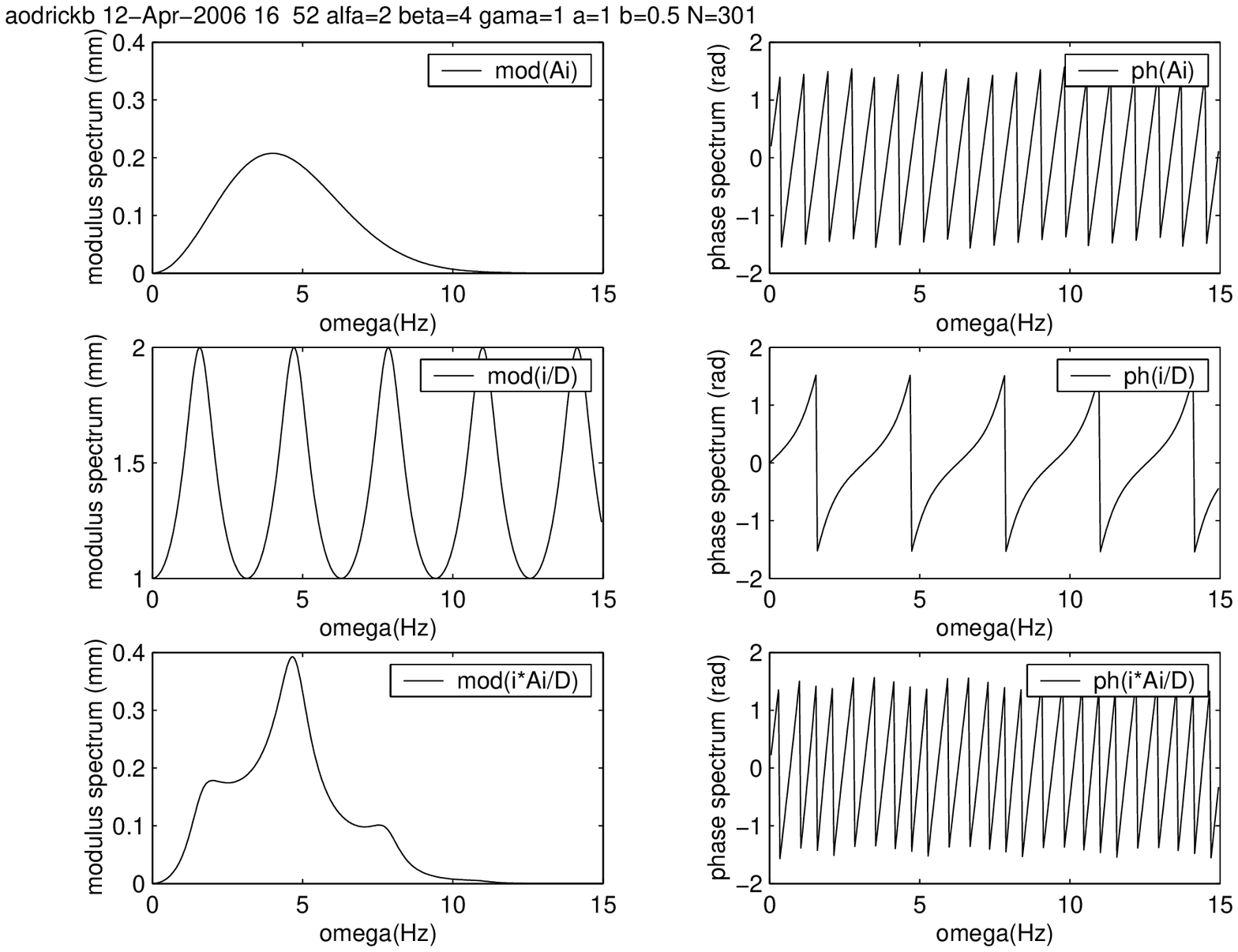}
  \caption{Spectrum of displacement response at $\mathbf{x}=(0,0)$.
  $\mathcal{A}=1$, $\alpha=2$, $\beta=4$, $\gamma=1$, $a=1$, $b=0.5$,
  corresponding to a case of merged pulses.  The left-hand
  curves
  pertain to moduli, and the right hand curves to phases of the spectra.}
  \label{aodrickb3}
  \end{center}
\end{figure}

\begin{figure}
[ptb]
\begin{center}
\includegraphics[scale=0.5] {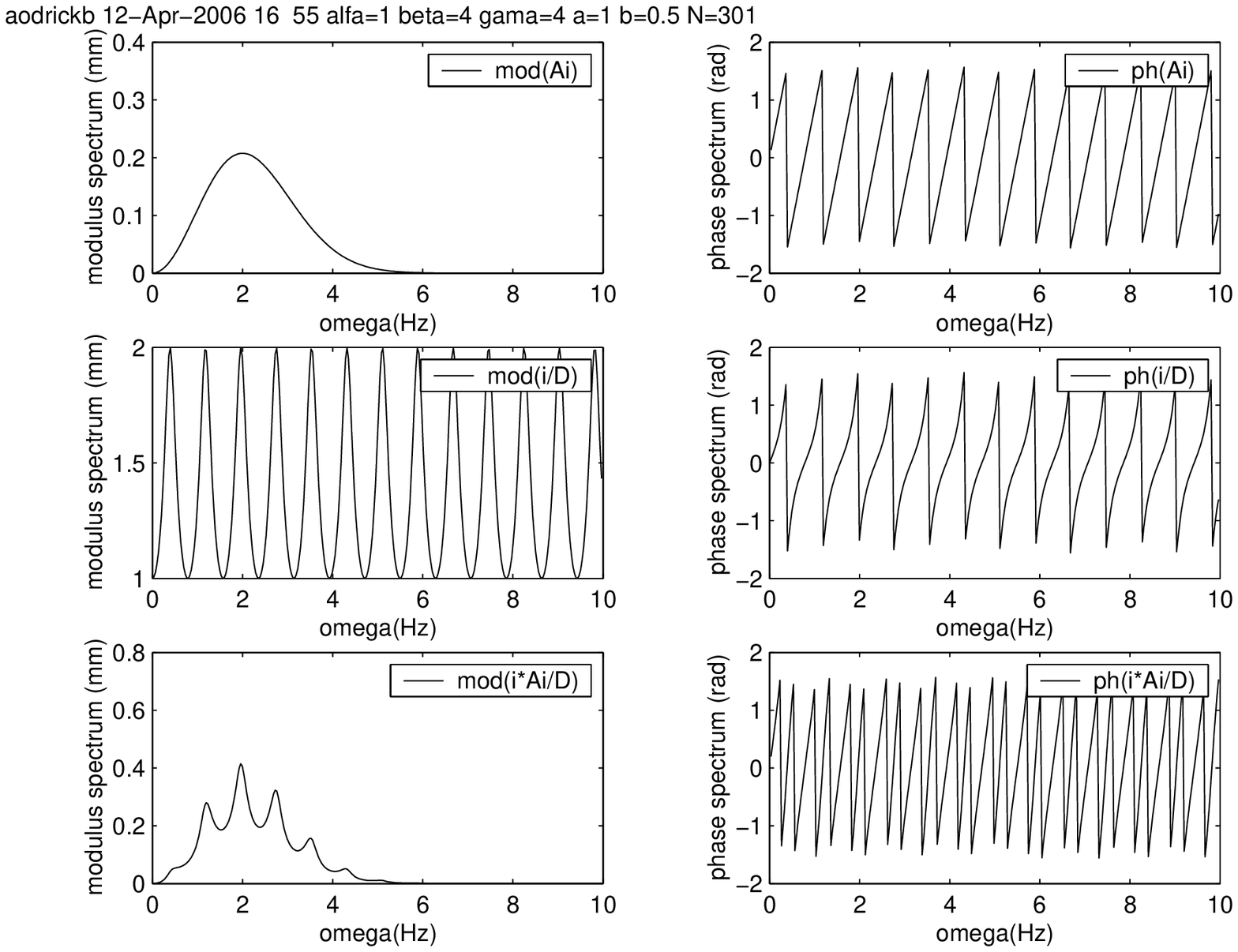}
  \caption{Spectrum of displacement response at $\mathbf{x}=(0,0)$.
  $\mathcal{A}=1$,
  $\alpha=1$, $\beta=4$, $\gamma=4$, $a=1$, $b=0.5$,
  corresponding to a case of separated pulses. The left-hand
  curves
  pertain to moduli, and the right hand curves to phases of the spectra.}
  \label{aodrickb5}
  \end{center}
\end{figure}

\begin{figure}
[ptb]
\begin{center}
\includegraphics[scale=0.5] {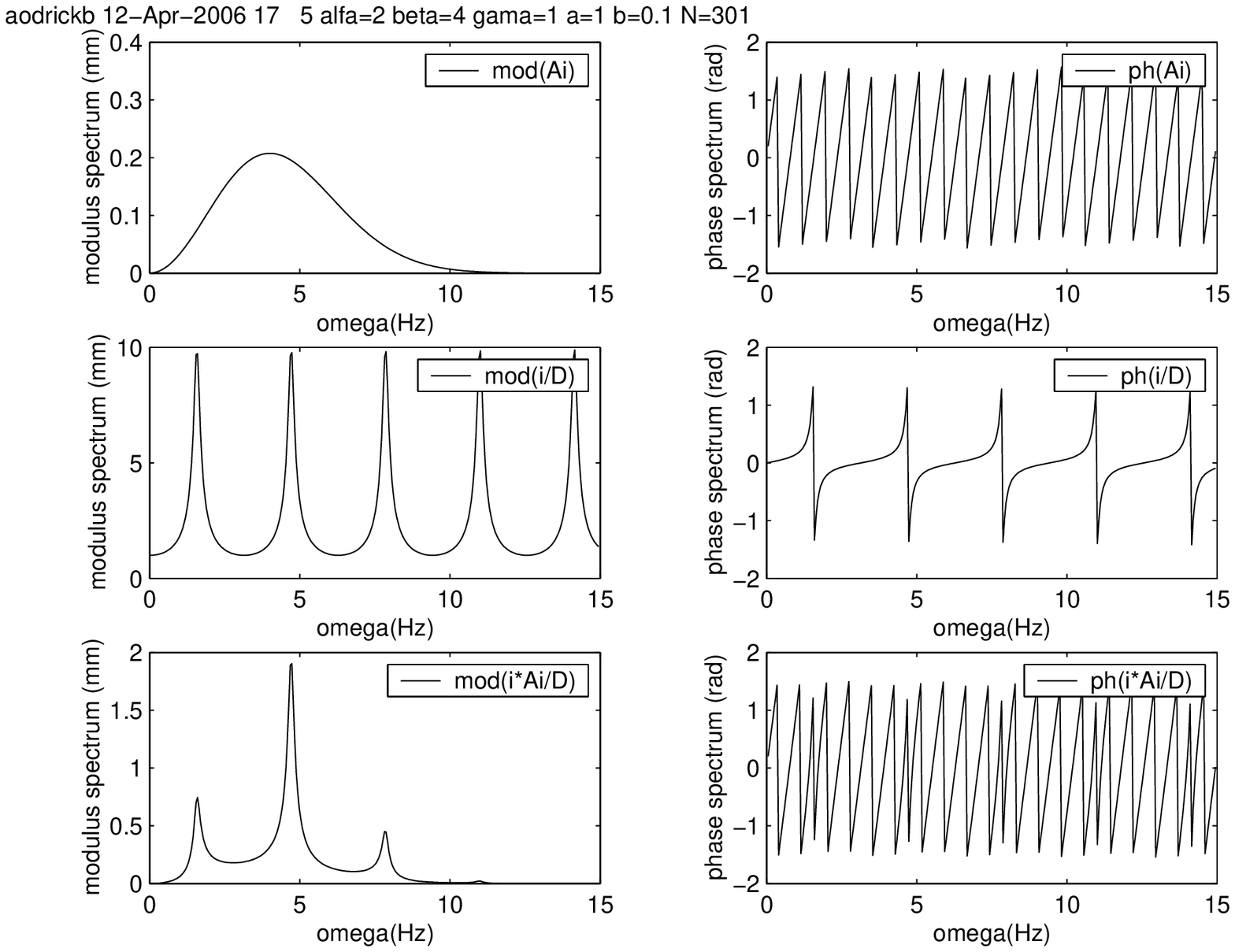}
  \caption{Spectrum of displacement response at $\mathbf{x}=(0,0)$.
  $\mathcal{A}=1$, $\alpha=2$, $\beta=4$, $\gamma=1$, $a=1$, $b=0.1$,
  corresponding to a case of merged pulses.  The left-hand
  curves
  pertain to moduli, and the right hand curves to phases of the spectra.}
  \label{aodrickb7}
  \end{center}
\end{figure}

\begin{figure}
[ptb]
\begin{center}
\includegraphics[scale=0.5] {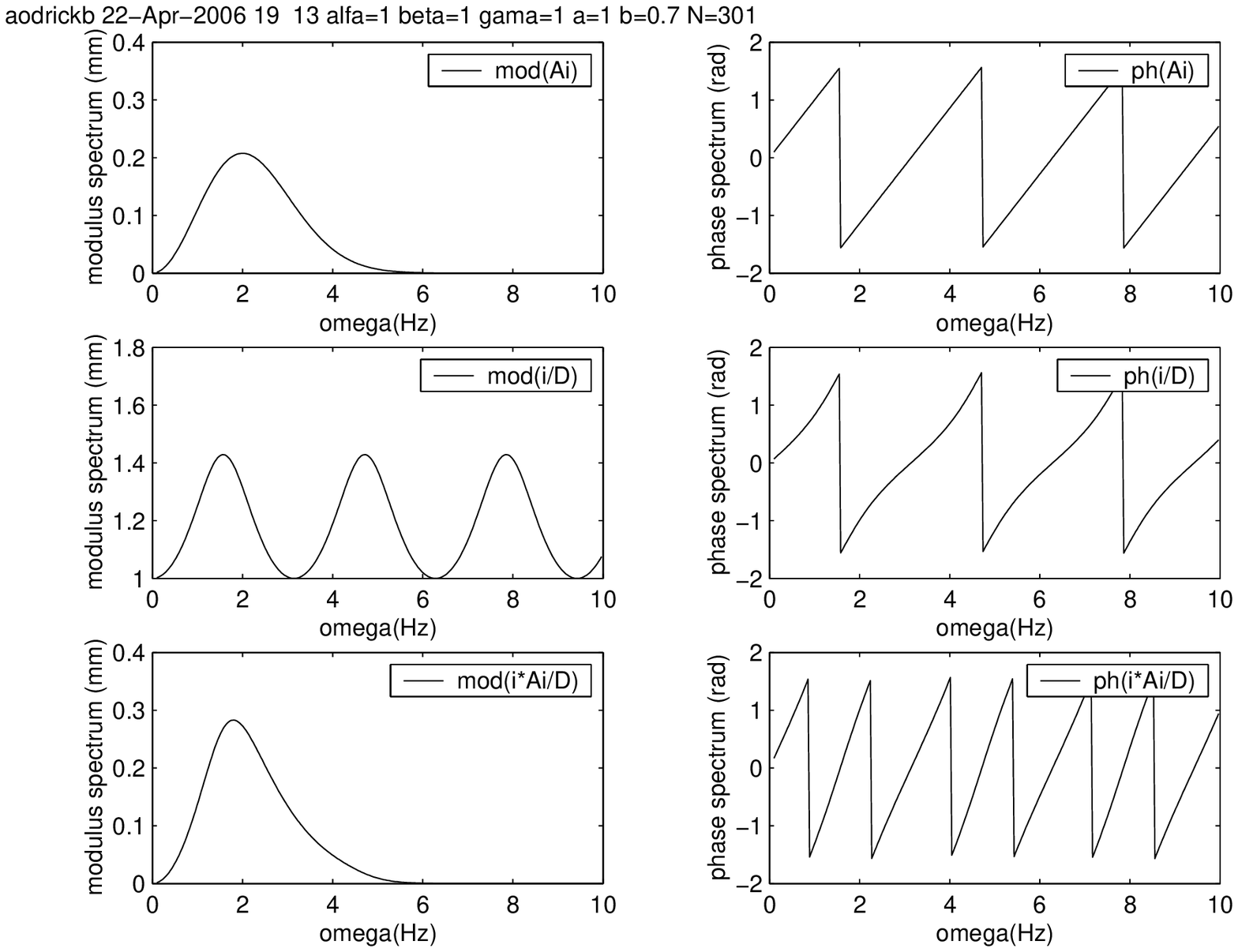}
  \caption{Spectrum of displacement response at $\mathbf{x}=(0,0)$.
  $\mathcal{A}=1$, $\alpha=1$, $\beta=1$, $\gamma=1$, $a=1$, $b=0.7$,
  corresponding to a case of a so-called "anomalous" pulse.  The left-hand
  curves
  pertain to moduli, and the right hand curves to phases of the spectra.}
  \label{aodrickb7a}
  \end{center}
\end{figure}

Recall that the frequency domain displacement response in the
layer is of the form
\begin{equation}\label{sevenhunseventyonea}
 u_{3}^{1}(\mathbf{x},\omega)=u_{3}^{1-}(\mathbf{x},\omega)+u_{3}^{1+}(\mathbf{x},\omega)~.
\end{equation}
with
\begin{equation}\label{sevenhunseventyoneb}
 u_{3}^{1\pm}(\mathbf{x},\omega)=
 \frac{\mathcal{R}^{1}(k_{1}^{i})}{2}A^{i-}(\omega)e^{i[k_{1}^{i}x_{1}\pm
 k_{2}^{1i}x_{2}]}~,
\end{equation}
so that one half of the total frequency domain displacement
response on the ground at point $\mathbf{x}=(0,0)$ is given by
\begin{equation}\label{sevenhunseventyonec}
 \frac{u_{3}^{1}(0,0,\omega)}{2}= u_{3}^{1\pm}(0,0,\omega)=
\frac{ \mathcal{R}^{1}(k_{1}^{i})}{2}A^{i-}(\omega)~,
\end{equation}
wherein:
\begin{equation}\label{sevenhunseventyoned}
 \frac{\mathcal{R}^{1}(k_{1}^{i})}{2}=\frac{i}{D(k_{1}^{i})}~,
 \end{equation}
\begin{equation}\label{sevenhunseventyonee}
 D(k_{1}^{i})=ia\cos(\gamma\omega)+b\sin(\gamma\omega)~,
 \end{equation}
\begin{equation}\label{sevenhunseventyonef}
a=1~~,~~b=b(\omega,s^{i})=
\frac{\mu^{1}(\omega)\kappa^{1i}(\omega,s^{i})}{\mu^{0}\kappa^{0i}(s^{i})}~~,
~~\gamma(\omega,s^{i})=\kappa^{1i}(\omega,s^{i})\frac{h}{c_{T}^{0}}~,
\end{equation}
\begin{equation}\label{sevenhunseventyoneg}
\kappa^{ji}:=\frac{k_{2}^{ji}}{k^{0}}
 =\sqrt{[c_{T}^{0}/c_{T}^{j}(\omega)]^{2}-(s^{i})^{2}}~.
\end{equation}
In the five figures \ref{aodrickb1}, \ref{aodrickb3},
\ref{aodrickb5}, \ref{aodrickb7}, \ref{aodrickb7a} we plot the
spectrum ( $A^{i-}(\omega)$) of a Ricker pulse excitation, the
transfer function $\frac{u_{3}^{1}(0,0,\omega)}{2A^{i}(\omega)}$,
and the spectrum of the displacement response
$\frac{u_{3}^{1}(0,0,\omega)}{2}$.
We note that the spectrum of displacement response  contains
spikes that are all the sharper and more intense the larger is the
contrast between $a$ and $b$. As will be shown further on, these
sharper spikes lead to larger-duration response in the time
domain.
\section{Time domain analysis of the reflection of a SH plane body
wave from a stress-free planar boundary overlying a soft layer
underlain by a hard substratum}\label{sec55}
%
\subsection{Obtention of the time domain response from the
frequency domain response}
We had
\begin{equation}\label{time1}
 u_{3}^{j}(\mathbf{x},t)=\int_{-\infty}^{\infty}u_{3}^{j}(\mathbf{x},\omega)\exp(-i\omega
  t)d\omega~~;~~j=0,1~,
\end{equation}
and due to the fact that $u_{3}^{j}(\mathbf{x},t)$ is a real
function, we must have
\begin{equation}\label{time2}
  [u_{3}^{j}(\mathbf{x},\omega)]^{*}=u_{3}^{j}(\mathbf{x},-\omega)~,
\end{equation}
(wherein the symbol $*$ designates the complex conjugate operator)
from which it follows that
\begin{equation}\label{time3}
  u_{3}^{j}(\mathbf{x},t)=2\Re\int_{0}^{\infty}u_{3}^{j}(\mathbf{x},\omega)\exp(-i\omega
  t)d\omega~.
\end{equation}
We shall employ (\ref{time3}) or (\ref{time1}) to obtain the
temporal response from the frequency response function
$u_{3}^{j}(\mathbf{x},\omega)$. More precisely, we shall be
concerned with the evaluation of
\begin{equation}\label{time4}
  u_{3}^{j\pm}(\mathbf{x},t)=\int_{-\infty}^{\infty}\left[ u_{3}^{j-}(\mathbf{x},\omega)+
  u_{3}^{j+}(\mathbf{x},\omega)\right] \exp(-i\omega
  t)d\omega~~;~~j=0,1~.
\end{equation}
\subsection{The frequency content and time history of a of a plane transient body wave}\label{sectdr1}
For a so-called {\it upward-propagating SH-plane wave}, we have
\begin{equation}\label{tdr116}
 u_{1}^{i}(\mathbf{x},\omega)=0~,~~
u_{2}^{i}(\mathbf{x},\omega)=0~,~~
u_{3}^{i}(\mathbf{x},\omega)=A^{i-}(\omega)\exp(i\mathbf{k}_{T}^{i-}(\omega)\cdot\mathbf{x})~.
\end{equation}
wherein
\begin{equation}\label{tdr116a}
\mathbf{k}_{T}^{i-}=(k_{T1}^{i},k_{T2}^{i-})~,~~k_{T1}^{i}:=k_{T}^{0}\sin\theta_{T}^{i}~,~~k_{T2}^{i\pm}:=\pm
k_{T}^{0}\cos\theta_{T}^{i}~,~~k_{T}^{0}(\omega)=\frac{\omega}{c_{T}(\omega)}
~,
\end{equation}
and $\theta_{T}^{i}$ is the incident angle measured clockwise from
the $+x_{2}$ axis. In the above relations, $A^{i-}(\omega)$ is
termed the {\it frequency-domain amplitude} or {\it spectrum} of
the incident SH plane wave.

The {\it time history} of this incident wave is
\begin{equation}\label{tdr117}
 \mathbf{u}^{i}(\mathbf{x},t)=\mathbf{u}^{0-}(\mathbf{x},t)=
 \int_{-\infty}^{\infty}\mathbf{u}^{0-}(\mathbf{x},\omega)\exp(-i\omega
 t)d\omega~,
\end{equation}
or
\begin{equation}\label{tdr118}
 \mathbf{u}^{i}(\mathbf{x},t)=\mathbf{u}^{0-}(\mathbf{x},t)=
 2\Re\int_{0}^{\infty}\mathbf{u}^{0-}(\mathbf{x},\omega)\exp(-i\omega t)d\omega~.
\end{equation}
\subsubsection{Spectrum and time history of a Ricker pulse}\label{stcricker}
\begin{figure}
[ptb]
\begin{center}
\includegraphics[scale=0.5] {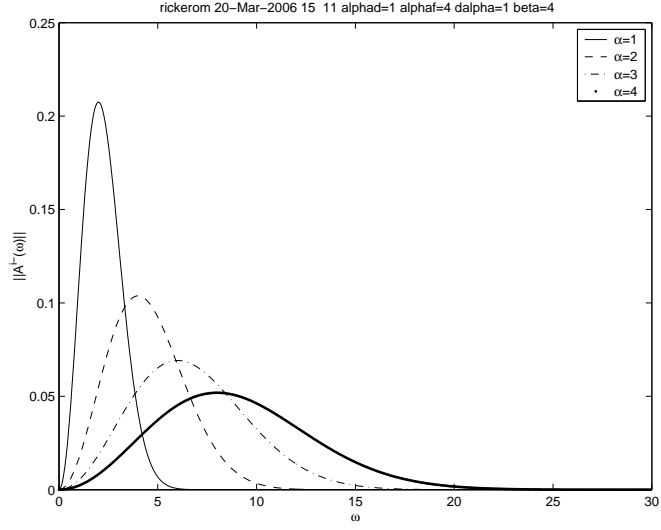}
  \caption{Modulus of the spectrum function $A^{i-}(\omega)$ versus
  $\omega$ for various values of $\alpha$. $\mathcal{A}=1$, $\beta=4$.}
  \label{rickerom}
  \end{center}
\end{figure}

\begin{figure}
[ptb]
\begin{center}
\includegraphics[scale=0.5] {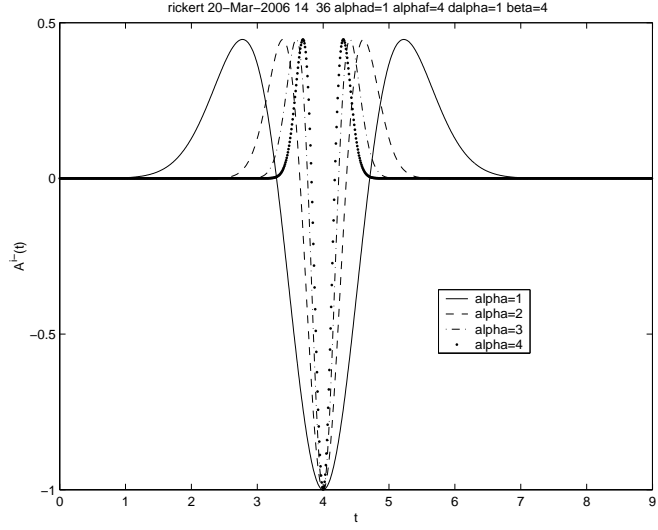}
  \caption{Time history  $A^{i-}(t)$ versus $t$ for various values of $\alpha$.
  $\beta=4$ and $\mathcal{A}=1$.}
  \label{rickert}
  \end{center}
\end{figure}

The amplitude spectrum $A^{i-}(\omega)$ of a Ricker pulse
(Sanchez-Sesma 1985)  is given by
\begin{equation}\label{tds133}
   A^{i-}(\omega)=-\mathcal{A}\frac{1}{\sqrt{\pi}}\frac{\omega^{2}}{4\alpha^{3}}
   \exp\left (i\beta\omega-\frac{\omega^{2}}{4\alpha^{2}}\right)~,
   \end{equation}
wherein $\mathcal{A}$, $\alpha$ and $\beta$ are  real constants
(i.e., independent of $\omega$). It follows that
\begin{equation}\label{tds134}
   [A^{i-}(\omega)]^{*}=-\mathcal{A}\frac{1}{\sqrt{\pi}}\frac{\omega^{2}}{4\alpha^{3}}
   \exp\left
   (-i\beta\omega-\frac{\omega^{2}}{4\alpha^{2}}\right)=A^{i-}(-\omega)~,
   \end{equation}
so that the temporal history associated with this pulse is
\begin{equation}\label{tds135}
   A^{i-}(t)=\int_{-\infty}^{\infty}A^{i-}(\omega)
\exp(-i\omega t)d\omega~.
\end{equation}
More precisely:
\begin{multline}\label{tds136}
   A^{i-}(t)=\left( -\mathcal{A}\frac{1}{4\alpha^{3}\sqrt{\pi}}\right) \int_{-\infty}^{\infty}\omega^{2}
\exp\left [i\omega(\beta-t)-\frac{\omega^{2}}{4\alpha^{2}}\right]
d\omega=
\\
\left( -\mathcal{A}\frac{1}{4\alpha^{3}\sqrt{\pi}}\right)
\int_{-\infty}^{\infty}\omega^{2} \exp\left
[-\frac{1}{4\alpha^{2}}\{\omega^{2}+4\alpha^{2}i\omega(t-\beta)\}\right]
d\omega~.
\end{multline}
But
\begin{equation}\label{tds137}
\omega^{2}+4\alpha^{2}i\omega(t-\beta)=[\omega+i2\alpha^{2}(t-\beta)]^{2}+4\alpha^{4}(t-\beta)^{2}
~,
\end{equation}
so that
\begin{equation}\label{tds138}
 A^{i-}(t)=\left( -\mathcal{A}\frac{1}{4\alpha^{3}\sqrt{\pi}}\right)\exp[-\alpha^{2}(t-\beta)^{2}]
 \int_{-\infty}^{\infty}\omega^{2}
 \exp\left[ -\frac{\{\omega+i2\alpha^{2}(t-\beta)\}^{2}}{4\alpha^{2}}\right] d\omega
~,
\end{equation}
which, after the change of variables
\begin{equation}\label{tds139}
 \varpi=\omega+i2\alpha^{2}(t-\beta)
~,
\end{equation}
becomes
\begin{multline}\label{tds140}
 A^{i-}(t)=\left( -\mathcal{A}\frac{1}{4\alpha^{3}\sqrt{\pi}}\right) \exp[-\alpha^{2}(t-\beta)^{2}]
 \int_{-\infty}^{\infty}[\varpi-i2\alpha^{2}(t-\beta)]^{2}
 \exp\left[ -\frac{\varpi^{2}}{4\alpha^{2}}\right] d\varpi:=
 \\
 \left( -\mathcal{A}\frac{1}{4\alpha^{3}\sqrt{\pi}}\right)
 \left[ J_{2}(t)-i4\alpha^{2}(t-\beta)J_{1}(t)-4\alpha^{4}(t-\beta)^{2}J_{0}(t)\right]
~,
\end{multline}
wherein:
\begin{equation}\label{tds141}
J_{0}(t)= \int_{-\infty}^{\infty}
 \exp\left[ -\frac{\varpi^{2}}{4\alpha^{2}}\right] d\varpi=2\int_{0}^{\infty}
 \exp\left[ -\frac{\varpi^{2}}{4\alpha^{2}}\right] d\varpi
~,
\end{equation}
\begin{equation}\label{tds142}
J_{1}(t)= \int_{-\infty}^{\infty}\varpi
 \exp\left[ -\frac{\varpi^{2}}{4\alpha^{2}}\right] d\varpi=0
~,
\end{equation}
\begin{equation}\label{tds143}
J_{2}(t)= \int_{-\infty}^{\infty}\varpi^{2}
 \exp\left[ -\frac{\varpi^{2}}{4\alpha^{2}}\right] d\varpi=2\int_{0}^{\infty}\varpi^{2}
 \exp\left[ -\frac{\varpi^{2}}{4\alpha^{2}}\right] d\varpi
~.
\end{equation}
Employing the following identities (Hodgman 1957):
\begin{equation}\label{tds144}
\int_{0}^{\infty} \exp \left( -\frac{\eta^{2}}{b^{2}}\right)
d\eta=\frac{b\sqrt{\pi}}{2}~,
\end{equation}
\begin{equation}\label{tds145}
\int_{0}^{\infty} \eta^{2}\exp \left(
-\frac{\eta^{2}}{b^{2}}\right) d\eta=\frac{b^{3}\sqrt{\pi}}{4}~,
\end{equation}
we finally obtain
\begin{equation}\label{tds146}
J_{0}(t)=2\alpha\sqrt{\pi}~~,~~J_{2}(t)=4\alpha^{3}\sqrt{\pi}~,
\end{equation}
so that
\begin{equation}\label{tds147}
A^{i-}(t)=\left(
-\mathcal{A}\frac{1}{4\alpha^{3}\sqrt{\pi}}\right) \left[
4\alpha^{3}\sqrt{\pi}-2\alpha\sqrt{\pi}4\alpha^{4}(t-\beta)^{2}\right]
\exp[-\alpha^{2}(t-\beta)^{2}]~,
\end{equation}
or
\begin{equation}\label{tds148}
A^{i-}(t)=\mathcal{A}\left[ -1+2\alpha^{2}(t-\beta)^{2}\right]
\exp[-\alpha^{2}(t-\beta)^{2}] ~.
\end{equation}
\newline
{\it Remark}
\newline
The maxima of $A^{i-}(t)/\mathcal{A}$ are obtained from
$dA^{i-}(t)/dt=0$, i.e.,
\begin{equation}\label{tds149}
2\alpha^{2}(t-\beta)\{2-[-1+2\alpha^{2}(t-\beta)^{2}]\}
\exp[-\alpha^{2}(t-\beta)^{2}]=0 ~.
\end{equation}
the solutions of which are
\begin{equation}\label{tds149a}
t_{0}=\beta~~,~~t_{-}=\beta-\sqrt{\frac{3}{2\alpha^{2}}}~~,
~~t_{-}=\beta+\sqrt{\frac{3}{2\alpha^{2}}}~.
\end{equation}
It follows that
\begin{equation}\label{tds150}
A^{i-}(t_{0})/\mathcal{A}=-1~~,~~A^{i-}(t_{\pm})/\mathcal{A}=2
\exp\left(-\frac{3}{2}\right) =0.4463 ~.
\end{equation}
\newline
{\it Remark}
\newline
Thus,  The  minimum of $A^{i-}(t)/\mathcal{A}$ is attained at
$t=t_{0}=\beta$ and is equal to $A^{i-}(t_{0})/\mathcal{A}=-1$.
This means that the minimum of $A^{i-}(t)/\mathcal{A}$ is
independent of both $\alpha$ and $\beta$. Furthermore, $\beta$ is
the instant at which the pulse attains its minimum, and this
minimum is also the maximum of $\|A^{i-}(t)/\mathcal{A}\|$.
\newline
\newline
{\it Remark}
\newline
$A^{i-}(t)/\mathcal{A}=0$ when $-1+2\alpha^{2}(t-\beta)^{2}=0$,
i.e., when $t=\beta\pm\frac{1}{2\alpha^{2}}$, so that $1/\alpha$
is an indicator of the width of the main lobe of the pulse, i.e.,
the larger is $\alpha$, the smaller is the width of the main lobe
of the pulse.
\newline
\newline
{\it Remark}
\newline
The maxima of $A^{i-}(t)/\mathcal{A}$ are attained at $t=t_{-}$
and $t=t_{+}$, and their value (0.4463) is independent of both
$\alpha$ and $\beta$.
\newline
\newline
{\it Remark}
\newline
The moduli of the spectra of the Ricker pulses for various values
of $\alpha$ are depicted in fig. \ref{rickerom}.
Note that these spectra do not depend on $\beta$. The latter only
affects the phase of $A^{i-}(\omega)$.
\newline
\newline
{\it Remark}
\newline
The time history of the Ricker pulses for various values of
$\alpha$ are depicted in fig. \ref{rickert} for the fixed value
$\beta=4$. This confirms the fact that the larger is $\alpha$, the
narrower is the Ricker pulse.
\newline
\newline
{\it Remark}
\newline
Let $G(t)$ be the Gaussian function
\begin{equation}\label{tds151}
G(t)=\frac{\mathcal{A}}{2\alpha^{2}}\exp[-\alpha
^{2}(t-\beta)^{2}]~.
\end{equation}
Then
\begin{equation}\label{tds152}
\frac{dG(t)}{dt}=-\mathcal{A}(t-\beta)\exp[-\alpha
^{2}(t-\beta)^{2}]~,~\frac{d^{2}G(t)}{dt^{2}}=\mathcal{A}[-1+2\alpha^{2}
(t-\beta)^{2}]\exp[-\alpha ^{2}(t-\beta)^{2}]=A^{i-}(t)~,
\end{equation}
which shows that the Ricker pulse is identical to the second time
derivative of the Gaussian pulse.
\subsubsection{Spectrum and time history of a Ricker pulse plane
body wave  propagating in free space}\label{secricker}
The wave we are concerned with is actually what was heretofore
termed the "incident plane wave". The plane body wave nature of
the disturbance is embodied in the frequency domain function
\begin{equation}\label{tds153}
u_{3}^{0-}(\mathbf{x},\omega)=
A^{i-}(\omega)\exp[i(k_{1}^{i}x_{1}-k_{2}^{0i}x_{2})]=A^{i-}(\omega)
\exp\left[ i\omega\left(
\frac{x_{1}}{c_{T}^{0}}s^{i}-\frac{x_{2}}{c_{T}^{0}}\kappa^{0i}\right)
\right] ~,
\end{equation}
wherein
\begin{equation}\label{tds154}
s^{i}=\sin\theta_{T}^{i}~~,~~\kappa^{0i}=\sqrt{1-(s^{i})^{2}}=
\cos\theta_{T}^{i}~.
\end{equation}
The Ricker pulse nature of the plane wave is embodied in the
spectrum function
\begin{equation}\label{tds154a}
   A^{i-}(\omega)=\left(
   -\mathcal{A}\frac{1}{\sqrt{\pi}}\frac{\omega^{2}}{4\alpha^{3}}\right)
   \exp\left (i\beta\omega-\frac{\omega^{2}}{4\alpha^{2}}\right)~,
   \end{equation}
so that the time history of this wave is given by the Fourier
transform
\begin{equation}\label{tds155}
u_{3}^{i}(\mathbf{x},t)=\int_{-\infty}^{\infty}
A^{i-}(\omega)\exp[- i\omega \tau(\mathbf{x},t,s^{i})]d\omega ~,
   \end{equation}
wherein
\begin{equation}\label{tds156}
\tau(\mathbf{x},t,s^{i})=t-\frac{x_{1}}{c_{T}^{0}}s^{i}+\frac{x_{2}}{c_{T}^{0}}c^{i}~.
   \end{equation}
We  assume that $M^{0}$ is non-dispersive, i.e., $\mu^{0}$ and
$c_{T}^{0}$ do not depend on $\omega$, so that
$\tau(\mathbf{x},t,s^{i})$ is independent of $\omega$. Then
$u_{3}^{i}(\mathbf{x},t)$ is of the same form as $A^{i-}(t)$ (see
(\ref{tds135})) if we replace therein $t$ by
$\tau(\mathbf{x},t,s^{i})$, i.e.,
\begin{equation}\label{tds157}
u_{3}^{i}(\mathbf{x},t)=A^{i-}(\tau(\mathbf{x}),t,s^{i})~,
   \end{equation}
so that employing the previous result in (\ref{tds148}):
\begin{equation}\label{tds158}
u_{3}^{i}(\mathbf{x},t)=\mathcal{A}\left[
-1+2\alpha^{2}\{\tau(\mathbf{x})-\beta\}^{2}\right]
\exp[-\alpha^{2}\{\tau(\mathbf{x},t,s^{i})-\beta\}^{2}] ~.
\end{equation}
{\it Remark}
\newline
The same remarks apply to this time history as to $A^{i-}(t)$
except that $\tau(\mathbf{x},t,s^{i})$ replaces $t$. Thus, the
Ricker pulse plane body wave attains its maximum (in absolute
value) at $\tau(\mathbf{x},t,s^{i})=\beta$, and the main lobe of
the pulse is all the narrower the larger is $\alpha$.
\subsection{Time history of the reflected and transmitted plane body wave
pulses in the basement and layer}
%
\subsubsection{Preliminaries}
Recall that $k^{j}=\frac{\omega}{c_{T}^{j}}$. We encountered
previously, in connection with the frequency domain response in
$\Omega_{0}$ and $\Omega_{1}$, plane wave functions of the type
\begin{multline}\label{th1}
u_{3}^{j\pm}(\mathbf{x},\omega):=\mathcal{R}^{j}(s^{i},\omega)A^{i-}(\omega)\exp[i(k_{1}^{i}x_{1}\pm
k_{2}^{ji}x_{2})]=
\\
\mathcal{R}^{j}(s^{i},\omega)A^{i-}(\omega)\exp\left[
i\omega\left( s^{i}\frac{x_{1}}{c_{T}^{0}}\pm
\kappa^{ji}\frac{x_{2}}{c_{T}^{0}}\right) \right] ~~;~~j=0,1~,
\end{multline}
wherein (recall that we assumed that $c_{T}^{0}$ does not depend
on $\omega$):
\begin{equation}\label{th1a}
\kappa^{ji}:=\frac{k_{2}^{ji}}{k^{0}}:=\frac{\sqrt{(k^{j})^{2}-(k_{1}^{i})^{2}}}{k^{0}}=
\sqrt{(k^{j}/k^{0})^{2}-(s^{i})^{2}}
 =\sqrt{[c_{T}^{0}/c_{T}^{j}(\omega)]^{2}-(s^{i})^{2}}~,
\end{equation}
and, recalling  that $\mu^{0}$  was also assumed not to depend on
$\omega$:
\begin{multline}\label{th2}
  \mathcal{R}^{0}(\omega,s^{i})=\left[
  \frac{
  \mu^{0}\kappa^{0i}(s^{i})
  \cos\left( \kappa^{1i}(\omega,s^{i})\frac{h}{c_{T}^{0}}\right) +
  i\mu^{1}(\omega)\kappa^{1i}(\omega,s^{i})
  \sin\left( \kappa^{1i}(\omega,s^{i})\frac{h}{c_{T}^{0}}\right)
  }
  {
  \mu^{0}\kappa^{0i}(s^{i})
  \cos\left( \kappa^{1i}(\omega,s^{i})\frac{h}{c_{T}^{0}}\right)-
  i\mu^{1}(\omega)\kappa^{1i}(\omega,s^{i})
  \sin\left( \kappa^{1i}(\omega,s^{i})\frac{h}{c_{T}^{0}}\right)
  }
  \right]
  e^{\left[ -2i\kappa^{0i}(s^{i})\frac{h}{c_{T}^{0}}\right] } =
  \\
  R^{0}(\omega,(s^{i})e^{\left[ -2i\kappa^{0i}(s^{i})\frac{h}{c_{T}^{0}}\right] }~,
\end{multline}
\begin{multline}\label{th3}
  \mathcal{R}^{1}(\omega,s^{i})=\left[
  \frac{
  2\mu^{0}\kappa^{0i}(s^{i})
  }
  {
  \mu^{0}\kappa^{0i}(s^{i})
  \cos\left( \kappa^{1i}(\omega,s^{i})\frac{h}{c_{T}^{0}}\right)-
  i\mu^{1}(\omega)\kappa^{1i}(\omega,s^{i})
  \sin\left( \kappa^{1i}(\omega,s^{i})\frac{h}{c_{T}^{0}}\right)
  }
  \right]
  e^{\left[ -i\kappa^{0i}(s^{i})\frac{h}{c_{T}^{0}}\right] }=
  \\
  R^{1}(\omega,(s^{i})e^{\left[ -i\kappa^{0i}(s^{i})\frac{h}{c_{T}^{0}}\right] }~,
\end{multline}
wherein
\begin{equation}\label{th3a}
R^{0}(\omega,s^{i})=\left[ \frac{ i \cos\left(
\kappa^{1i}(\omega,s^{i})\frac{h}{c_{T}^{0}}\right)
  -\frac{\mu^{1}(\omega)\kappa^{1i}(\omega,s^{i})}{\mu^{0}\kappa^{0i}(s^{i})}
  \sin\left( \kappa^{1i}(\omega,s^{i})\frac{h}{c_{T}^{0}}\right) }
  {i\cos\left( \kappa^{1i}(\omega,s^{i})\frac{h}{c_{T}^{0}}\right) +
  \frac{\mu^{1}(\omega)\kappa^{1i}(\omega,s^{i})}{\mu^{0}\kappa^{0i}(s^{i})}
  \sin\left( \kappa^{1i}(\omega,s^{i})\frac{h}{c_{T}^{0}}\right) }\right]
\end{equation}
\begin{equation}\label{th3b}
  R^{1}(\omega,s^{i})=\left[ \frac{2i}
  {i\cos(\kappa^{1i}(\omega,s^{i})k^{0}h)+
  \frac{\mu^{1}(\omega)\kappa^{1i}(\omega,s^{i})}{\mu^{0}\kappa^{0i}(s^{i})}
  \sin(\kappa^{1i}(\omega,s^{i})\frac{h}{c_{T}^{0}})}\right]~.
\end{equation}
Thus, we are faced with the problems of evaluating the integrals
\begin{equation}\label{th3c}
  u_{3}^{0+}(\mathbf{x},t)=\int_{-\infty}^{\infty}A^{i-}(\omega)R^{0}(\omega,s^{i})
  \exp\left[ i\omega\left( s^{i}\frac{x_{1}}{c_{T}^{0}}+\kappa^{0i}(s^{i})\frac{x_{2}}{c_{T}^{0}}-
  2\kappa^{0i}(s^{i})\frac{h}{c_{T}^{0}}-t\right) \right] d\omega~,
\end{equation}
\begin{equation}\label{th3d}
  u_{3}^{1\pm}(\mathbf{x},t)=\int_{-\infty}^{\infty}A^{i-}(\omega)\frac{R^{1}(\omega,s^{i})}{2}
  \exp\left[ i\omega\left( s^{i}\frac{x_{1}}{c_{T}^{0}}\pm\kappa^{1i}(\omega,s^{i})\frac{x_{2}}{c_{T}^{0}}-
  \kappa^{0i}(s^{i})\frac{h}{c_{T}^{0}}-t\right) \right] d\omega~,
\end{equation}
or
\begin{equation}\label{th3e}
  u_{3}^{0+}(\mathbf{x},t)=\int_{-\infty}^{\infty}A^{i-}(\omega)R^{0}(\omega,s^{i})
  \exp\left[ -i\omega\tau^{0+}(\mathbf{x},t,s^{i})\right] d\omega~,
\end{equation}
\begin{equation}\label{th3f}
  u_{3}^{1\pm}(\mathbf{x},t)=\int_{-\infty}^{\infty}A^{i-}(\omega)\frac{R^{1}(\omega,s^{i})}{2}
  \exp\left[ -i\omega\tau^{1\pm}(\mathbf{x},t,\omega,s^{i})\right] d\omega~,
\end{equation}
wherein:
\begin{equation}\label{th3g}
 \tau^{0+}(\mathbf{x},t,s^{i}):=
 t-s^{i}\frac{x_{1}}{c_{T}^{0}}-\kappa^{0i}(s^{i})\frac{x_{2}}{c_{T}^{0}}+
  2\kappa^{0i}(s^{i})\frac{h}{c_{T}^{0}}~,
\end{equation}
\begin{equation}\label{th3h}
 \tau^{1\pm}(\mathbf{x},t,\omega,s^{i}):=
 t-s^{i}\frac{x_{1}}{c_{T}^{0}}\mp\kappa^{1i}(\omega,s^{i})\frac{x_{2}}{c_{T}^{0}}+
  \kappa^{0i}(s^{i})\frac{h}{c_{T}^{0}}~.
\end{equation}
\newline
In the following, whenever numerical results are given, they will
apply to the field on the ground at point $\mathbf{x}=(0,0)$.

Recall that the time domain displacement response in the layer is
of the form
\begin{equation}\label{th3i}
 u_{3}^{1}(\mathbf{x},t)=u_{3}^{1-}(\mathbf{x},t)+u_{3}^{1+}(\mathbf{x},t)~.
\end{equation}
so that
\begin{equation}\label{th3j}
 \frac{u_{3}^{1}(0,0,t)}{2}= u_{3}^{1\pm}(0,0,t)~.
\end{equation}
It is this function, related to the temporal displacement response
on the ground, which will be depicted in the graphs that are
presented hereafter.
\subsubsection{Evaluation of $u_{3}^{1\pm}(\mathbf{x},t)$ for Ricker pulse excitation
by a rectangle quadrature scheme}
The time history of the displacement  field in $\Omega_{1}$ is of
the form
\begin{multline}\label{rq1}
  u_{3}^{1\pm}(\mathbf{x},t)=\int_{-\infty}^{\infty}A^{i-}(\omega)\frac{R^{1}(\omega,s^{i})}{2}
  \exp\left[ -i\omega\tau^{1\pm}(\mathbf{x},t,\omega,s^{i})\right] d\omega=
  \\
  i\int_{-\infty}^{\infty}\frac{A^{i-}(\omega)}{D(\omega,s^{i})}
  \exp\left[ -i\omega\tau^{1\pm}(\mathbf{x},t,\omega,s^{i})\right] d\omega~,
\end{multline}
wherein
\begin{equation}\label{rq2}
 D(\omega,s^{i})=ia\cos(\gamma\omega)+b\sin(\gamma\omega)
 ~,
\end{equation}
\begin{equation}\label{rq3}
a=1~~,~~b=b(\omega,s^{i})=
\frac{\mu^{1}(\omega)\kappa^{1i}(\omega,s^{i})}{\mu^{0}\kappa^{0i}(s^{i})}~~,
~~\gamma(\omega,s^{i})=\kappa^{1i}(\omega,s^{i})\frac{h}{c_{T}^{0}}
 ~,
\end{equation}
and the Ricker pulse excitation is represented by
\begin{equation}\label{rq4}
A^{i-}(\omega)=\left(
-\frac{\mathcal{A}}{4\alpha^{3}\sqrt{\pi}}\right)
\omega^{2}\exp\left(
i\beta\omega-\frac{\omega^{2}}{4\alpha^{2}}\right)
 ~.
\end{equation}
Consequently
\begin{equation}\label{rq5}
A^{i-}(-\omega)=\left(
-\frac{\mathcal{A}}{4\alpha^{3}\sqrt{\pi}}\right)
\omega^{2}\exp\left(
-i\beta\omega-\frac{\omega^{2}}{4\alpha^{2}}\right)
 ~,
\end{equation}
\begin{equation}\label{rq6}
 D(-\omega,s^{i})=ia\cos(\gamma\omega)-b\sin(\gamma\omega)
 ~,
\end{equation}
whence
\begin{multline}\label{rq8}
u_{3}^{1\pm}(\mathbf{x},t)=\left(
-\frac{i\mathcal{A}}{4\alpha^{3}\sqrt{\pi}}\right)\int_{0}^{\infty}\omega^{2}\Big[
 \frac{\exp\left\{ -i[\beta-\tau^{1\pm}(\mathbf{x},t,\omega,s^{i})]\omega\right\} }
 {ia\cos(\gamma\omega)-b\sin(\gamma\omega)} +
 \\
 \frac{\exp\left\{ i[\beta-\tau^{1\pm}(\mathbf{x},t,\omega,s^{i})]\omega\right\} }
 {ia\cos(\gamma\omega)+b\sin(\gamma\omega)}
\Big]
  \exp\left[ -\frac{\omega^{2}}{4\alpha^{2}}\right] d\omega~.
\end{multline}
This formula shows that the integrand is an
exponentially-decreasing function of $\omega$ so that the
numerical evaluation of the integral should pose no problems. In
particular, we replace the upper limit of the integral by
$\omega_{max}$, where the latter is such that $\exp\left[
-\frac{\omega_{max}^{2}}{4\alpha^{2}}\right] <<1$, so that we are
faced with the computation of
\begin{multline}\label{rq9}
u_{3}^{1\pm}(\mathbf{x},t)=\left(
-\frac{i\mathcal{A}}{4\alpha^{3}\sqrt{\pi}}\right)\int_{0}^{\omega_{max}}\omega^{2}\Big[
 \frac{\exp\left\{ -i[\beta-\tau^{1\pm}(\mathbf{x},t,\omega,s^{i})]\omega\right\} }
 {ia\cos(\gamma(\omega,s^{i})\omega)-b(\omega,s^{i})\sin(\gamma(\omega,s^{i})\omega)} +
 \\
 \frac{\exp\left\{ i[\beta-\tau^{1\pm}(\mathbf{x},t,\omega,s^{i})]\omega\right\} }
 {ia\cos(\gamma(\omega,s^{i})\omega)+b(\omega,s^{i})\sin(\gamma(\omega,s^{i})\omega)}
\Big]
  \exp\left[ -\frac{\omega^{2}}{4\alpha^{2}}\right] d\omega~.
\end{multline}
To this end, we therefore employ the simplest method: rectangle
quadrature. We divide the interval $[0,\omega_{max}]$ into $L$
equal sub-intervals, of width $\delta=\omega_{max}/L$, and
centered at points $\omega_{l}=(2l-1)\delta/2~;~l=1,2,...,L$, so
that
\begin{multline}\label{rq10}
u_{3}^{1\pm}(\mathbf{x},t)\approx\left(
-\frac{i\mathcal{A}\delta}{4\alpha^{3}\sqrt{\pi}}\right)\sum_{l=1}^{L}\omega_{l}^{2}\Big[
 \frac{\exp\left\{ -i[\beta-\tau^{1\pm}(\mathbf{x},t,\omega_{l},s^{i})]\omega\right\} }
 {ia\cos(\gamma(\omega_{l},s^{i})\omega_{l})-b(\omega_{l},s^{i})\sin(\gamma(\omega_{l},s^{i})\omega_{l})} +
 \\
 \frac{\exp\left\{ i[\beta-\tau^{1\pm}(\mathbf{x},t,\omega_{l},s^{i})]\omega\right\} }
 {ia\cos(\gamma(\omega_{l},s^{i})\omega)+b(\omega_{l},s^{i})\sin(\gamma(\omega_{l},s^{i})\omega)}
\Big]
  \exp\left[ -\frac{\omega_{l}^{2}}{4\alpha^{2}}\right] d\omega~.
\end{multline}
This result seems to indicate that $u_{3}^{1\pm}(\mathbf{x},t)$ is
complex. However, combining into one the two terms in $[~]$, we
find
\begin{multline}\label{rq11}
u_{3}^{1\pm}(\mathbf{x},t)\approx\left(
\frac{-\mathcal{A}\delta}{4\alpha^{3}\sqrt{\pi}}\right)
\sum_{l=1}^{L}\left(
 \frac{\omega_{l}^{2}}
 {a^{2}\cos^{2}\left( \gamma(\omega_{l},s^{i})\omega_{l}\right )+b^{2}(\omega_{l},s^{i})
 \sin^{2}\left( \gamma(\omega_{l},s^{i}\omega\right) }\right) \times
 \\
 \Big[ \left( a+b(\omega_{l},s^{i})\right) \cos\left\{\gamma(\omega_{l},s^{i})+
 \beta-\tau^{1\pm}(\mathbf{x},t,\omega_{l},s^{i})\omega_{l}\right\}
 +
 \\
 \left( a-b(\omega_{l},s^{i})\right)
 \cos\left\{\gamma(\omega_{l},s^{i})-
 \beta+\tau^{1\pm}(\mathbf{x},t,\omega_{l},s^{i})\omega_{l}\right\}
 \Big]
 \exp\left[ -\frac{\omega_{l}^{2}}{4\alpha^{2}}\right] d\omega
 ~,
\end{multline}
which shows that $u_{3}^{1\pm}(\mathbf{x},t)$ is indeed real (at
least for real $\gamma(\omega_{l},s^{i})$).

Eq. (\ref{rq11}) is the formula we employ for the numerical
evaluation of $u_{3}^{1\pm}(\mathbf{x},t)$. Note that this formula
is exact in the limits $\omega_{max}\rightarrow\infty$,
$L\rightarrow\infty$.
\subsubsection{Evaluation of $u_{3}^{1\pm}(\mathbf{x},t)$ for Ricker pulse excitation
by a power series quadrature scheme}
Once again, the time history of the displacement  field in
$\Omega_{1}$ is of the form
\begin{multline}\label{psq1}
  u_{3}^{1\pm}(\mathbf{x},t)=\int_{-\infty}^{\infty}A^{i-}(\omega)\frac{R^{1}(\omega,s^{i})}{2}
  \exp\left[ -i\omega\tau^{1\pm}(\mathbf{x},t,\omega,s^{i})\right] d\omega=
  \\
  i\int_{-\infty}^{\infty}\frac{A^{i-}(\omega)}{D(\omega,s^{i})}
  \exp\left[ -i\omega\tau^{1\pm}(\mathbf{x},t,\omega,s^{i})\right] d\omega~,
\end{multline}
wherein
\begin{equation}\label{psq2}
 D(\omega,s^{i})=ia\cos(\gamma\omega)+b\sin(\gamma\omega)
 ~,
\end{equation}
\begin{equation}\label{psq3}
a=1~~,~~b=b(\omega,s^{i})=
\frac{\mu^{1}(\omega)\kappa^{1i}(\omega,s^{i})}{\mu^{0}\kappa^{0i}(s^{i})}~~,
~~\gamma(\omega,s^{i})=\kappa^{1i}(\omega,s^{i})\frac{h}{c_{T}^{0}}
 ~,
\end{equation}
and the Ricker pulse excitation is represented by
\begin{equation}\label{psq4}
A^{i-}(\omega)=\left(
-\frac{\mathcal{A}}{4\alpha^{3}\sqrt{\pi}}\right)
\omega^{2}\exp\left(
i\beta\omega-\frac{\omega^{2}}{4\alpha^{2}}\right)
 ~.
\end{equation}
From here on, we adopt a strategy that is different from the one
in the previous section, notably by writing $D(\omega,s^{i})$ as
\begin{equation}\label{psq4a}
 D(\omega,s^{i})=\frac{i}{2}\left[ \{a-b(\omega,s^{i})\}e^{i\gamma(\omega,s^{i})\omega}+
\{a+b(\omega,s^{i})\}e^{-i\gamma(\omega,s^{i})\omega}\right]
 ~.
\end{equation}
The idea is to express $D$ as something like $1-\chi$ and then to
employ the power series expansion of $(1-\chi)^{-1}$ to express
$D^{-1}$, but we have to be careful to have $\|\chi\|<1$ in order
for this series to converge. Thus, we write
\begin{multline}\label{psq5}
 D(\omega,s^{i}):=D^{-}(\omega,s^{i})=-\frac{i}{2}\{b(\omega,s^{i})-a\}
 e^{i\gamma(\omega,s^{i})\omega}\left[ 1-\left( \frac{b(\omega,s^{i})+a}{b(\omega,s^{i})-a}\right)
 e^{-2i\gamma(\omega,s^{i})\omega}\right]
 ~~;
 \\
 ~~\left\| \left( \frac{b(\omega,s^{i})+a}{b(\omega,s^{i})-a}\right)  e^{-2i\gamma(\omega,s^{i})\omega}\right\| <1~,
\end{multline}
\begin{multline}\label{psq6}
 D(\omega,s^{i}):=D^{+}(\omega,s^{i})=\frac{i}{2}\{b(\omega,s^{i})+a\}
 e^{-i\gamma(\omega,s^{i})\omega}\left[ 1-\left( \frac{b(\omega,s^{i})-a}{b(\omega,s^{i})+a}\right)
 e^{2i\gamma(\omega,s^{i})\omega}\right]
 ~~;
 \\
 ~~\left\| \left( \frac{b(\omega,s^{i})-a}{b(\omega,s^{i})+a}\right)  e^{2i\gamma(\omega,s^{i})\omega}\right\| <1~,
\end{multline}
so that:
\begin{equation}\label{psq7}
 [D^{-}(\omega,s^{i})]^{-1}=\left(
 -\frac{i}{2}\{b(\omega,s^{i})-a\} e^{i\gamma(\omega,s^{i})\omega}\right)^{-1}\sum_{m=0}^{\infty}
\left( \frac{b(\omega,s^{i})+a}{b(\omega,s^{i})-a}\right)^{m}
 e^{-2im\gamma(\omega,s^{i})\omega}
 ~,
\end{equation}
\begin{equation}\label{psq8}
 [D^{+}(\omega,s^{i})]^{-1}=\left(
 \frac{i}{2}\{b(\omega,s^{i})+a\} e^{-i\gamma(\omega,s^{i})\omega}\right)^{-1}\sum_{m=0}^{\infty}
\left( \frac{b(\omega,s^{i})-a}{b(\omega,s^{i})+a}\right)^{m}
 e^{2im\gamma(\omega,s^{i})\omega}
 ~,
\end{equation}
or
\begin{equation}\label{psq9}
 i[D^{-}(\omega,s^{i})]^{-1}=\sum_{m=0}^{\infty}C_{m}^{-}(\omega,s^{i})
 e^{-i(2m+1)\gamma(\omega,s^{i})\omega}
 ~,
\end{equation}
\begin{equation}\label{psq10}
 i[D^{+}(\omega,s^{i})]^{-1}=\sum_{m=0}^{\infty}C_{m}^{+}(\omega,s^{i})
 e^{i(2m+1)\gamma(\omega,s^{i})\omega}
 ~,
\end{equation}
wherein
\begin{equation}\label{psq11}
  C_{m}^{-}(\omega,s^{i})=-2
 \frac{\left( b(\omega,s^{i})+a\right)^{m}}{\left(
 b(\omega,s^{i})-a\right) ^{m+1}}
 ~,
\end{equation}
\begin{equation}\label{psq12}
  C_{m}^{+}(\omega,s^{i})=2
 \frac{\left( b(\omega,s^{i})-a\right)^{m}}{\left(
 b(\omega,s^{i})+a\right) ^{m+1}}
 ~.
\end{equation}
At this point we recall that
\begin{equation}\label{psq13}
a=i~~,~~b=b(\omega,s^{i})=
\frac{\mu^{1}(\omega)\kappa^{1i}(\omega,s^{i})}{\mu^{0}\kappa^{0i}(s^{i})}~,
~~\gamma(\omega,s^{i})=\kappa^{1i}(\omega,s^{i})\frac{h}{c_{T}^{0}}~
,~~\kappa^{1i}(\omega,s^{i})=\sqrt{\left(
\frac{c_{T}^{0}}{c_{T}^{1}(\omega)}\right)^{2}-(s^{i})^{2}}
 ~,
\end{equation}
\begin{equation}\label{psq13a}
 \tau^{0+}(\mathbf{x},t,s^{i}):=
 t-s^{i}\frac{x_{1}}{c_{T}^{0}}-\kappa^{0i}(s^{i})\frac{x_{2}}{c_{T}^{0}}+
  2\kappa^{0i}(s^{i})\frac{h}{c_{T}^{0}}~,
\end{equation}
\begin{equation}\label{psq13b}
 \tau^{1\pm}(\mathbf{x},t,\omega,s^{i}):=
 t-s^{i}\frac{x_{1}}{c_{T}^{0}}\mp\kappa^{1i}(\omega,s^{i})\frac{x_{2}}{c_{T}^{0}}+
  \kappa^{0i}(s^{i})\frac{h}{c_{T}^{0}}~.
\end{equation}
and {\it assume henceforth that $\mu^{1}$ and $c_{T}^{1}$ do not
depend on $\omega$} so that $\kappa^{1i}=\kappa^{1i}(s^{i})$ and
\begin{equation}\label{psq14}
a=i~~,~~b=b(s^{i})=
\frac{\mu^{1}\kappa^{1i}(s^{i})}{\mu^{0}\kappa^{0i}(s^{i})}~~,
~~\gamma(s^{i})=\kappa^{1i}(s^{i})\frac{h}{c_{T}^{0}}
 ~,
\end{equation}
\begin{equation}\label{psq14a}
 \tau^{1\pm}(\mathbf{x},t,s^{i})=
 t-s^{i}\frac{x_{1}}{c_{T}^{0}}\mp\kappa^{1i}(s^{i})\frac{x_{2}}{c_{T}^{0}}+
  \kappa^{0i}(s^{i})\frac{h}{c_{T}^{0}}~.
\end{equation}
whence
\begin{equation}\label{psq15}
  C_{m}^{-}(s^{i})=-2
 \frac{\left( b(s^{i})+a\right)^{m}}{\left(
 b(s^{i})-a\right) ^{m+1}}
 ~,
\end{equation}
\begin{equation}\label{psq16}
  C_{m}^{+}(s^{i})=2
 \frac{\left( b(s^{i})-a\right)^{m}}{\left(
 b(s^{i})+a\right) ^{m+1}}
 ~.
\end{equation}
Furthermore, we assume that $c_{T}^{0}>c_{T}^{1}$ so that
$\kappa^{1i}$ is real for all $|s^{i}|\leq 1$. Consequently
$\gamma(\omega,s^{i})$ is real for all $|s^{i}|\leq 1$, and the
condition $\left\| \left( \frac{b(s^{i})+a}{
 b(s^{i})-a}\right) \exp[-2i\gamma(s^{i},\omega)]\right\| \gtrless1$ reduces to $\left\|\frac{b(s^{i})+a}{
 b(s^{i})-a}\right\|\gtrless1$.

It follows that the time history of the field in $\Omega_{1}$
takes the form
\begin{multline}\label{psq17}
  u_{3}^{1\pm}(\mathbf{x},t)=
  i\int_{-\infty}^{\infty}\frac{A^{i-}(\omega)}{D(\omega,s^{i})}
  \exp\left[ -i\omega\tau^{1\pm}(\mathbf{x},t,s^{i})\right] d\omega=
  \\
  i\sum_{m=0}^{\infty} C_{m}^{-}(s^{i})\int_{-\infty}^{\infty}A^{i-}(\omega)
  \exp\left[ -i\omega\{\tau^{1\pm}(\mathbf{x},t,s^{i})+(2m+1)\gamma(s^{i})\}\right] d\omega~~;
  \\
  ~~\left\| \left( \frac{b(s^{i})+a}{b(s^{i})-a} \right) \right\| <1~,
\end{multline}
\begin{multline}\label{psq18}
  u_{3}^{1\pm}(\mathbf{x},t)=
  i\int_{-\infty}^{\infty}\frac{A^{i-}(\omega)}{D(\omega,s^{i})}
  \exp\left[ -i\omega\tau^{1\pm}(\mathbf{x},t,s^{i})\right] d\omega=
  \\
  i\sum_{m=0}^{\infty} C_{m}^{+}(s^{i})\int_{-\infty}^{\infty}A^{i-}(\omega)
 \exp\left[ -i\omega\{\tau^{1\pm}(\mathbf{x},t,s^{i})-(2m+1)\gamma(s^{i})\}\right] d\omega~~;
 \\
  ~~\left\| \left( \frac{b(s^{i})+a}{b(s^{i})-a}\right) \right\| >1~,
\end{multline}
or, on account of the Ricker pulse nature of the excitation,
\begin{multline}\label{psq19}
  u_{3}^{1\pm}(\mathbf{x},t)=
  \\
  \left(
\frac{-\mathcal{A}}{4\alpha^{3}\sqrt{\pi}}\right)
  \sum_{m=0}^{\infty} C_{m}^{-}(s^{i})\int_{-\infty}^{\infty}\omega^{2}
  \exp\left[ i\omega\{\beta-\tau^{1\pm}(\mathbf{x},t,s^{i})-
  (2m+1)\gamma(s^{i})\}-\frac{\omega^{2}}{4\alpha^{2}}\right] d\omega~~;
  \\
  ~~\left\| \left( \frac{b(s^{i})+a}{b(s^{i})-a}\right) \right\| <1~,
\end{multline}
\begin{multline}\label{psq20}
  u_{3}^{1\pm}(\mathbf{x},t)=
  \\
\left( \frac{-\mathcal{A}}{4\alpha^{3}\sqrt{\pi}}\right)
\sum_{m=0}^{\infty}
C_{m}^{+}(s^{i})\int_{-\infty}^{\infty}\omega^{2}
 \exp\left[ i\omega\{\beta-\tau^{1\pm}(\mathbf{x},t,s^{i})+
 (2m+1)\gamma(s^{i})\}-\frac{\omega^{2}}{4\alpha^{2}}\right] d\omega~~;
 \\
  ~~\left\| \left( \frac{b(s^{i})+a}{b(s^{i})-a}\right) \right\|
  >1~.
\end{multline}
We recall here the previous result
\begin{multline}\label{psq21}
   A^{i-}(t)=\left( -\mathcal{A}\frac{1}{4\alpha^{3}\sqrt{\pi}}\right) \int_{-\infty}^{\infty}\omega^{2}
\exp\left [i\omega(\beta-t)-\frac{\omega^{2}}{4\alpha^{2}}\right]
d\omega=
\\
\mathcal{A}\left[ -1+2\alpha^{2}(t-\beta)^{2}\right]
\exp[-\alpha^{2}(t-\beta)^{2}] ~,
\end{multline}
so that
\begin{multline}\label{psq22}
  u_{3}^{1\pm}(\mathbf{x},t)=\mathcal{A}
  \sum_{m=0}^{\infty} C_{m}^{-}(s^{i})\left[ -1+
  2\alpha^{2}\left( \tau^{1\pm}(\mathbf{x},t,s^{i})-\beta_{m}^{-}(s^{i})\right)
  ^{2}\right] \times
  \\
  \exp\left[ -\alpha^{2}\left( \tau^{1\pm}(\mathbf{x},t,\omega,s^{i})-\beta_{m}^{-}(s^{i})\right) ^{2}\right] ~~;
  ~~\left\| \left( \frac{b(s^{i})+a}{b(s^{i})-a}\right) \right\| <1~,
\end{multline}
\begin{multline}\label{psq23}
  u_{3}^{1\pm}(\mathbf{x},t)=\mathcal{A}
\sum_{m=0}^{\infty}C_{m}^{+}(s^{i})\left[ -1+
  2\alpha^{2}\left( \tau^{1\pm}(\mathbf{x},t,s^{i})-\beta_{m}^{+}(s^{i})\right)
  ^{2}\right] \times
  \\
  \exp\left[ -\alpha^{2}\left( \tau^{1\pm}(\mathbf{x},t,\omega,s^{i})-\beta_{m}^{+}(s^{i})\right) ^{2}\right] ~~;
  ~~\left\| \left( \frac{b(s^{i})+a}{b(s^{i})-a}\right) \right\|
  >1~,
\end{multline}
wherein
\begin{equation}\label{psq24}
\beta_{m}^{\pm}(s^{i}):=\beta\pm(2m+1)\gamma(s^{i})~.
\end{equation}
Although these formulae are exact, they are not suitable for
computation due to the presence of the infinite series therein.
Actually, for practical (numerical) purposes, we limit the series
to a finite ($M+1$) number of terms (which is justified by the
fact that the terms of the series are exponentially-decreasing
with $m$) so that
\begin{multline}\label{psq25}
  u_{3}^{1\pm}(\mathbf{x},t)\approx\mathcal{A}
  \sum_{m=0}^{M} C_{m}^{-}(s^{i})\left[ -1+
  2\alpha^{2}\left( \tau^{1\pm}(\mathbf{x},t,s^{i})-\beta_{m}^{-}(s^{i})\right)
  ^{2}\right] \times
  \\
  \exp\left[ -\alpha^{2}\left( \tau^{1\pm}(\mathbf{x},t,\omega,s^{i})-\beta_{m}^{-}(s^{i})\right) ^{2}\right] ~~;
  ~~\left\| \left( \frac{b(s^{i})+a}{b(s^{i})-a}\right)\right\| <1~,
\end{multline}
\begin{multline}\label{psq26}
  u_{3}^{1\pm}(\mathbf{x},t)\approx\mathcal{A}
\sum_{m=0}^{M}C_{m}^{+}(s^{i})\left[ -1+
  2\alpha^{2}\left( \tau^{1\pm}(\mathbf{x},t,s^{i})-\beta_{m}^{+}(s^{i})\right)
  ^{2}\right] \times
  \\
  \exp\left[ -\alpha^{2}\left( \tau^{1\pm}(\mathbf{x},t,\omega,s^{i})-\beta_{m}^{+}(s^{i})\right) ^{2}\right] ~~;
  ~~\left\| \left( \frac{b(s^{i})+a}{b(s^{i})-a}\right) \right\|
  >1~,
\end{multline}
These last two formulae (which are exact in the limit
$M\rightarrow\infty$) form the basis of what we term the {\it
power series quadrature method} for the computation of the time
history of the displacement field in $\Omega_{1}$.
\subsubsection{Evaluation of $u_{3}^{1\pm}(\mathbf{x},t)$ for Ricker pulse excitation
by the complex frequency pole-residue convolution
scheme}\label{cfprc}
Once again, the time history of the displacement field in
$\Omega_{1}$ is of the form
\begin{multline}\label{cfprc1}
  u_{3}^{1\pm}(\mathbf{x},t)=\int_{-\infty}^{\infty}A^{i-}(\omega)\frac{R^{1}(\omega,s^{i}}{2}
  \exp\left[ -i\omega\tau^{1\pm}(\mathbf{x},t,\omega,s^{i})\right] d\omega=
  \\
  i\int_{-\infty}^{\infty}\frac{A^{i-}(\omega)}{D(\omega,s^{i})}
  \exp\left[ -i\omega\tau^{1\pm}(\mathbf{x},t,\omega,s^{i})\right] d\omega~,
\end{multline}
wherein
\begin{equation}\label{cfprc2}
 D(\omega,s^{i})=ia\cos(\gamma\omega)+b\sin(\gamma\omega)
 ~,
\end{equation}
\begin{multline}\label{cfprc3}
a=1~~,~~b=b(\omega,s^{i})=
\frac{\mu^{1}(\omega)\kappa^{1i}(\omega,s^{i})}{\mu^{0}\kappa^{0i}(s^{i})}~~,
~~\gamma(\omega,s^{i})=\kappa^{1i}(\omega,s^{i})\frac{h}{c_{T}^{0}}~~,
\\
~~ \tau^{1\pm}(\mathbf{x},t,\omega,s^{i}):=
 t-s^{i}\frac{x_{1}}{c_{T}^{0}}\mp\kappa^{1i}(\omega,s^{i})\frac{x_{2}}{c_{T}^{0}}+
  \kappa^{0i}(s^{i})\frac{h}{c_{T}^{0}}
 ~,
\end{multline}
and the Ricker pulse excitation is represented by
\begin{equation}\label{cfprc4}
A^{i-}(\omega)=\left(
-\frac{\mathcal{A}}{4\alpha^{3}\sqrt{\pi}}\right)
\omega^{2}\exp\left(
i\beta\omega-\frac{\omega^{2}}{4\alpha^{2}}\right)
 ~.
\end{equation}
Before going into details, we recall some general considerations.
Eqs. (\ref{cfprc1})-(\ref{cfprc4}) show that the task is to
evaluate the Fourier integral of a product of two functions
$F_{1}$ and $F_{2}$:
\begin{equation}\label{cfprc5}
  u(\tau)=\int_{-\infty}^{\infty}F_{1}(\omega)F_{2}(\omega)
  \exp(-i\omega\tau) d\omega~.
\end{equation}
We make use of the Fourier integral representations of
$F_{1}(\omega)$ and $F_{2}(\omega)$:
\begin{equation}\label{cfprc6}
F_{1}(\omega)=\frac{1}{2\pi}\int_{-\infty}^{\infty}F_{1}(t_{1})
  \exp(i\omega t_{1}) dt_{1}~~,~~F_{2}(\omega)=\frac{1}{2\pi}\int_{-\infty}^{\infty}F_{2}(t_{2})
  \exp(i\omega t_{2}) dt_{2}~,
\end{equation}
whose inverses are:
\begin{equation}\label{cfprc6a}
F_{1}(t_{1})=\int_{-\infty}^{\infty}F_{1}(\omega)
  \exp(-i\omega t_{1}) d\omega~~,~~F_{2}(t_{2})=\int_{-\infty}^{\infty}F_{2}(\omega)
  \exp(-i\omega t_{2}) d\omega~.
\end{equation}
We introduce (\ref{cfprc6} into (\ref{cfprc5}) so as to obtain:
\begin{equation}\label{cfprc7}
  u(\tau)=\left( \frac{1}{2\pi}\right) ^{2}\int_{-\infty}^{\infty}dt_{1}F_{1}(t_{1})
  \int_{-\infty}^{\infty}dt_{2}F_{2}(t_{2})
  \int_{-\infty}^{\infty}d\omega\exp\left[ -i\omega(\tau-t_{1}-t_{2}))\right] ~,
\end{equation}
or, due to the fact that
\begin{equation}\label{cfprc8}
  \int_{-\infty}^{\infty}d\omega\exp\left[ -i\omega(\tau-t_{1}-t_{2})\right] d\omega=
  2\pi\delta(\tau-t_{1}-t_{2})~,
\end{equation}
we find
\begin{equation}\label{cfprc9}
  u(\tau)= \frac{1}{2\pi}\int_{-\infty}^{\infty}F_{1}(t_{1})
  F_{2}(\tau-t_{1})dt_{1}
  ~,
\end{equation}
which is a {\it convolution} integral.

We can go a step further by expressing everything for positive
times:
\begin{equation}\label{cfprc10}
  u(\tau)= \frac{1}{2\pi}\int_{0}^{\infty}\left[ F_{1}(-t_{1})
  F_{2}(\tau+t_{1})+F_{1}(t_{1})
  F_{2}(\tau-t_{1})\right] dt_{1}
  ~.
\end{equation}
We now apply this formula to evaluate the time history of
response. As usual, we assume that {\it the media are
dispersionless}, so that
$\tau^{1\pm}(\mathbf{x},t,\omega,s^{i})=\tau^{1\pm}(\mathbf{x},t,s^{i})$,
whence
\begin{equation}\label{cfprc11}
  u_{3}^{1\pm}(\mathbf{x},t) = \frac{1}{2\pi}\int_{0}^{\infty}\left[ F_{1}(-t_{1})
  F_{2}\left( \tau^{1\pm}(\mathbf{x},t,s^{i})+t_{1}\right) +F_{1}(t_{1})
  F_{2}\left( \tau^{1\pm}(\mathbf{x},t,s^{i})-t_{1}\right) \right] dt_{1}
  ~.
\end{equation}
We choose
\begin{equation}\label{cfprc12}
 F_{2}(\omega)=A^{i-}(\omega)~~,~~ F_{1}(\omega)=\frac{i}{D(\omega,s^{i})}
  ~.
\end{equation}
We recall the result of (\ref{tds148})
\begin{equation}\label{cfprc13}
 F_{2}(t_{2})=A^{i-}(t_{2})=\mathcal{A}\left[ -1+2\alpha^{2}(t_{2}-\beta)^{2}\right]
\exp[-\alpha^{2}(t_{2}-\beta)^{2}] ~.
\end{equation}
and the Fourier inverse of $F_{1}(\omega)$ is
\begin{equation}\label{cfprc14}
 F_{1}(t_{1})=\int_{-\infty}^{\infty}\frac{i}{D(\omega,s^{i})}
  \exp(-i\omega t_{1}) d\omega~.
\end{equation}
Assuming as before that $b$ is real, we note immediately that
although the denominator of the integral in (\ref{cfprc14}) does
not vanish for {it real} $\omega$, it can vanish for {\it complex}
$\omega$. This suggests that the integral can be evaluated by use
of the Cauchy theorem by appealing to a suitable integration path
in the complex $\omega$ plane. Actually $D$ vanishes at an
infinite number of locations in the  complex
$\omega=\omega^{'}+\omega^{"}$ plane, so that we prefer to proceed
as following.

In order to stress the fact that the integration variable in
(\ref{cfprc14}) is real (i.e., $\omega^{'}$), and for other
reasons, we re-write the integral as
\begin{equation}\label{cfpr5}
  F_{1}(t)=\int_{-\infty}^{\infty}\frac{i}{D(\omega',s^{i})}
  \exp(-i\omega t_{}) d\omega'~. ~,
\end{equation}
wherein
\begin{equation}\label{cfpr8}
  D(\omega^{'},s^{i})=ia\cos(\gamma\omega^{'})+b\sin(\gamma\omega^{'})
  ~.
\end{equation}
We assume that $D$  vanishes for a denumerable set of {\it
complex} frequencies $\{\omega_{m}~;~m\in\mathbb{Z}\}$, i.e.,
\begin{equation}\label{cfpr14}
D(\omega_{m},s^{i})=0~~;~~m\in\mathbb{Z}
  ~.
\end{equation}
This suggests expanding $D(\omega,s^{i})$ in a Taylor series
around $\omega=\omega_{m}$:
\begin{equation}\label{cfpr15}
D(\omega,s^{i})=D(\omega_{m},s^{i})+
(\omega-\omega_{m})\dot{D}(\omega_{m},s^{i})+...
  ~,
\end{equation}
wherein
\begin{equation}\label{cfpr16}
\dot{D}(s^{i},\omega_{m})=\frac{\partial
D(\omega,s^{i})}{\partial\omega}\Big|
_{\omega=\omega_{m}}=\dot{D}(\omega_{m},s^{i})
  ~.
\end{equation}
On account of (\ref{cfpr14}) we have
\begin{equation}\label{cfpr17}
D(\omega,s^{i})\approx
(\omega-\omega_{m})\dot{D}(s^{i},\omega_{m})
  ~,
\end{equation}
whence
\begin{equation}\label{cfpr18}
\frac{1}{D(\omega',s^{i})}\approx
\frac{1}{(\omega'-\omega_{m})\dot{D}(\omega_{m},s^{i})}
  ~.
\end{equation}
Now let us turn to the issue of the actual locations of the zeros
of $D$. We search for the complex roots $\omega$ of
\begin{equation}\label{cfpr19}
D(\omega,s^{i})=ia\cos(\gamma\omega)+b\sin(\gamma\omega)=
ia\cos(\gamma(\omega^{'}+i\omega^{"}))+b\sin(\gamma(\omega^{'}+i\omega^{"})=0
  ~,
\end{equation}
and assume, as was implicit (or explicitly stated), that $a$, $b$
and $\gamma$ are real. Then
\begin{equation}\label{cfpr20}
D(\omega,s^{i})=\sin(\gamma\omega^{'})\left[
a\sinh(\gamma\omega^{"}) +b\cosh(\gamma\omega^{"})\right]
+i\cos(\gamma\omega^{'})\left[
a\cosh(\gamma\omega^{"})+b\sinh(\gamma\omega^{"})\right] =0
  ~,
\end{equation}
or, owing to the fact that we have a mixture of real and complex
quantitites:
\begin{equation}\label{cfpr21}
\sin(\gamma\omega^{'})\left[ a\sinh(\gamma\omega^{"})
+b\cosh(\gamma\omega^{"})\right]=0
  ~,
\end{equation}
\begin{equation}\label{cfpr22}
\cos(\gamma\omega^{'})\left[
a\cosh(\gamma\omega^{"})+b\sinh(\gamma\omega^{"})\right] =0
  ~.
\end{equation}
Thus, we have two families of solutions, the first of which
correspond to:
\begin{equation}\label{cfpr23}
\sin(\gamma\omega^{'})=0~~,~~a\cosh(\gamma\omega^{"})+b\sinh(\gamma\omega^{"})
=0~,
\end{equation}
and the second of which correspond to:
\begin{equation}\label{cfpr24}
\cos(\gamma\omega^{'})=0~~,~~ a\sinh(\gamma\omega^{"})
+b\cosh(\gamma\omega^{"})=0
  ~.
\end{equation}
The (so-called {\it even}) solutions of the first family are:
\begin{equation}\label{cfpr25}
\omega_{m}^{e'}=\frac{m\pi}{\gamma}~~;~~m\in\mathbb{Z}~,
\end{equation}
\begin{equation}\label{cfpr26}
\omega_{m}^{e"}=\frac{1}{2\gamma}\ln\left(\frac{b-a}{b+a}\right)~,
\end{equation}
and the (so-called {\it odd}) solutions of the second family are:
\begin{equation}\label{cfpr27}
\omega_{m}^{o'}=\frac{(2m+1)\pi}{2\gamma}~~;~~m\in\mathbb{Z}~,
\end{equation}
\begin{equation}\label{cfpr28}
\omega_{m}^{o"}=\frac{1}{2\gamma}\ln\left(\frac{a-b}{a+b}\right)~.
\end{equation}
{\it Remark}
\newline
The real parts of $\omega_{m}$ are independent of both $a$ and
$b$.
\newline
\newline
{\it Remark}
\newline
The imaginary parts of the even frequencies are independent of
$m$.
\newline
\newline
{\it Remark}
\newline
The imaginary parts of the odd frequencies are independent of $m$.
\newline
\newline
{\it Remark}
\newline
Since the $\ln(~)$ functions in these formulae are supposed to be
real, the even solutions apply only when $\frac{b-a}{b+a}>0$ and
the odd solutions apply only when  $\frac{b-a}{b+a}<0$.
\newline
\newline
{\it Remark}
\newline
Owing to the facts that: i) $\gamma>0$, ii)
$0<\frac{b-a}{b+a}=1-\frac{2a}{b+a}<1$ we have (for
$\frac{b-a}{b+a}>0$)
\begin{equation}\label{cfpr29}
\omega^{e"}=\frac{1}{2\gamma}\ln\left( \frac{b-a}{b+a}\right) <0~.
\end{equation}
{\it Remark}
\newline
Owing to the facts that: i) $\gamma>0$, ii)
$0<\frac{a-b}{a+b}=1-\frac{2b}{a+b}<1$ we have (for
$\frac{a-b}{a+b}>0$)
\begin{equation}\label{cfpr30}
\omega^{o"}=\frac{1}{2\gamma}\ln\left( \frac{a-b}{a+b}\right) <0~.
\end{equation}

Let us now evaluate $\dot{D}$. We have
\begin{multline}\label{cfpr31}
\dot{D}(\omega_{m})=\gamma\left[
-ia\sin(\gamma\omega_{m})+b\cos(\gamma\omega_{m}) \right]
=\gamma\left[ -ia \sin\left(
\gamma(\omega_{m}^{'}+i\omega_{m}^{"})\right) +b\cos\left(
\gamma(\omega_{m}^{'}+i\omega_{m}^{"})\right) \right] =
\\
\gamma\cos(\gamma\omega_{m}^{'})\left[
a\sinh(\gamma\omega_{m}^{"})+b\cosh(\gamma\omega_{m}^{"})\right]
+i\gamma\sin(\gamma\omega_{m}^{'})\left[
-a\cosh(\gamma\omega_{m}^{"})-b\sinh(\gamma\omega_{m}^{"})\right]
~.
\end{multline}
Then:
\begin{multline}\label{cfpr32}
\dot{D}(\omega_{m}^{e})=
\\
\gamma\cos(\gamma\omega_{m}^{e'})\left[
a\sinh(\gamma\omega_{m}^{e"})+b\cosh(\gamma\omega_{m}^{e"})\right]
+i\gamma\sin(\gamma\omega_{m}^{e'})\left[
-a\cosh(\gamma\omega_{m}^{e"})-b\sinh(\gamma\omega_{m}^{e"})\right]
=
\\
\gamma\cos(\gamma\omega_{m}^{e'})\left[
a\sinh(\gamma\omega_{m}^{e"})+b\cosh(\gamma\omega_{m}^{e"})\right]
=-\gamma(-1)^{m}\left( \frac{b^{2}-a^{2}}{a}\right)
\sinh(\gamma\omega_{m}^{e"}) ~,
\end{multline}
\begin{multline}\label{cfpr33}
\dot{D}(\omega_{m}^{o})=
\\
\gamma\cos(\gamma\omega_{m}^{o'})\left[
a\sinh(\gamma\omega_{m}^{o"})+b\cosh(\gamma\omega_{m}^{o"})\right]
+i\gamma\sin(\gamma\omega_{m}^{o'})\left[
-a\cosh(\gamma\omega_{m}^{o"})-b\sinh(\gamma\omega_{m}^{o"})\right]
=
\\
\gamma\sin(\gamma\omega_{m}^{o'})\left[
-a\cosh(\gamma\omega_{m}^{o"})-b\sinh(\gamma\omega_{m}^{o"})\right]
=i\gamma(-1)^{m}\left( \frac{b^{2}-a^{2}}{a}\right)
\cosh(\gamma\omega_{m}^{o"}) ~.
\end{multline}
The question that arises is whether we have to deal with either
even or odd solutions. Recall that
\begin{equation}\label{cfpr34}
a=1~~,~~b(\omega,s^{i})=\frac{\mu^{1}(\omega)\kappa^{1i}(\omega,s^{i})}{\mu^{0}\kappa^{0i}(\omega,s^{i})}~~,
~~\gamma(\omega,s^{i})=\kappa^{1i}(\omega,s^{i})\frac{h}{c_{T}^{0}}~,
\end{equation}
and, under the (previous) assumption of a dispersionless material
in the layer,
\begin{equation}\label{cfpr35}
a=1~~,~~b(s^{i})=\frac{\mu^{1}\kappa^{1i}(s^{i})}{\mu^{0}\kappa^{0i}(s^{i})}~~,~~
\gamma(s^{i})=\kappa^{1i}(s^{i})\frac{h}{c_{T}^{0}}~,
\end{equation}
so that
\begin{equation}\label{cfpr36}
b^{2}\gtrless a^{2}~~\Leftrightarrow~~ (\mu^{1})^{2}\left[ \left(
\frac{c_{T}^{0}}{c_{T}^{1}}\right) ^{2} -(s^{i}) ^{2}\right]
\gtrless (\mu^{0}) ^{2}\left[ 1-(s^{i}) ^{2}\right] ~,
\end{equation}
or, recalling that $ (c_{T}^{j}) ^{2}=\frac{\mu^{j}}{\rho^{j}}$,
$b^{2}\gtrless a^{2}$ is equivalent:
\begin{equation}\label{cfpr37}
\frac{\mu^{1}\rho^{1}}{\mu^{0}\rho^{0}}- \left(
\frac{\mu^{1}}{\mu^{0}}s^{i}\right) ^{2} \gtrless \left[ 1-(s^{i})
^{2}\right] ~.
\end{equation}
Recall that $0\leq|s^{i}|\leq 1$, so that for $s^{i}=0$ (i.e.,
normal incidence), $b^{2}\gtrless a^{2}$ (or $b\gtrless a$) is
equivalent to
\begin{equation}\label{cfpr38}
\mu^{1}\rho^{1}\gtrless\mu^{0}\rho^{0}~.
\end{equation}
whereas for  $s^{i}=\pm 1$ (i.e., grazing incidence),
$b^{2}\gtrless a^{2}$ (or $b\gtrless a$) is equivalent to
\begin{equation}\label{cfpr39}
\rho^{1}\gtrless\rho^{0}~.
\end{equation}
In the (geophysical) case of interest herein, i.e., dealing with a
soft layer overlying a hard substratum, we have $\mu^{0}>\mu^{1}$
and $\rho^{0}>\rho^{1}$, so that we are clearly in the situation
$b<a$ for all incidence angles. {\it This situation is that of odd
solutions}.

\begin{figure}
[ptb]
\begin{center}
\includegraphics[scale=0.5] {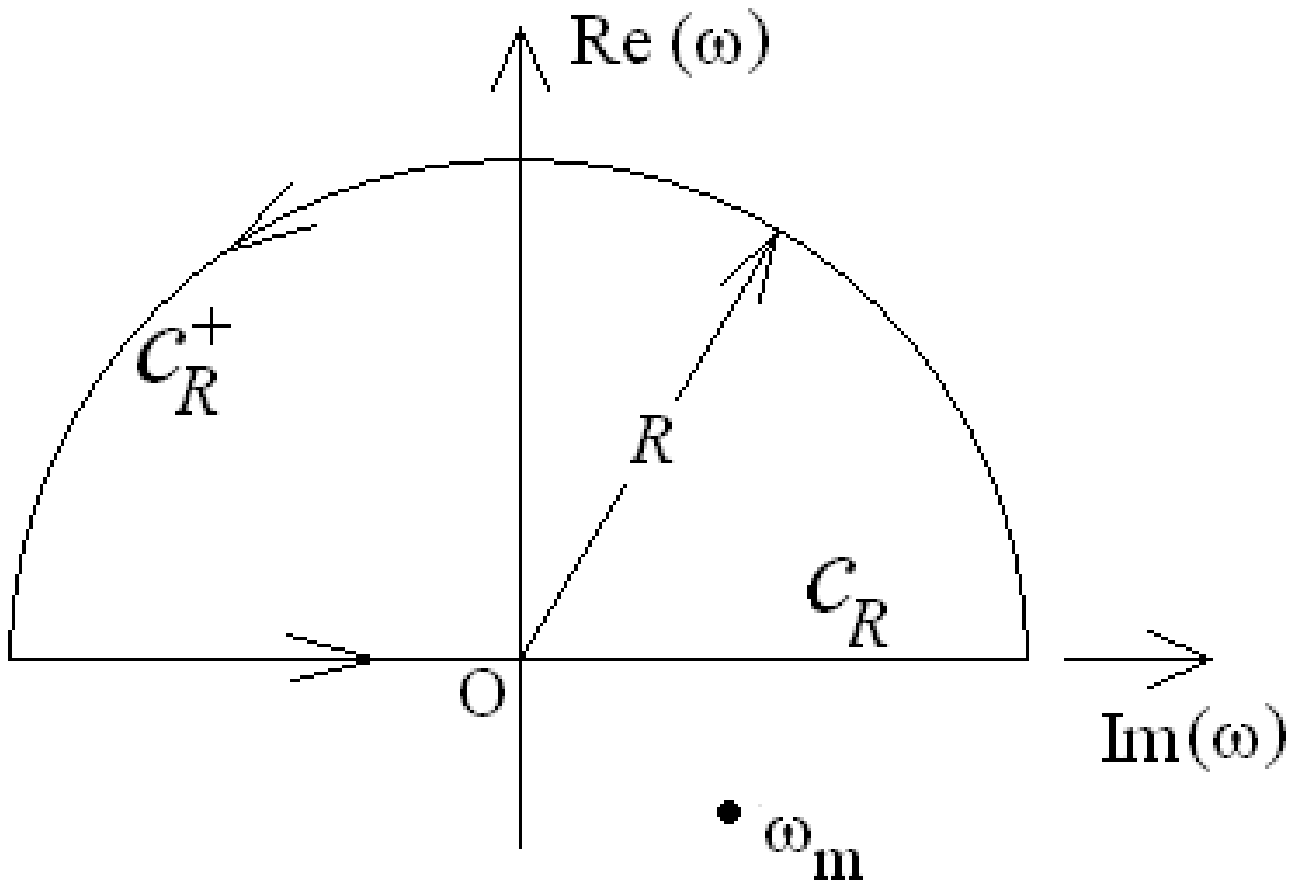}
  \caption{Contour $\mathcal{C}^{+}$.}
  \label{cplus}
  \end{center}
\end{figure}

\begin{figure}
[ptb]
\begin{center}
\includegraphics[scale=0.5] {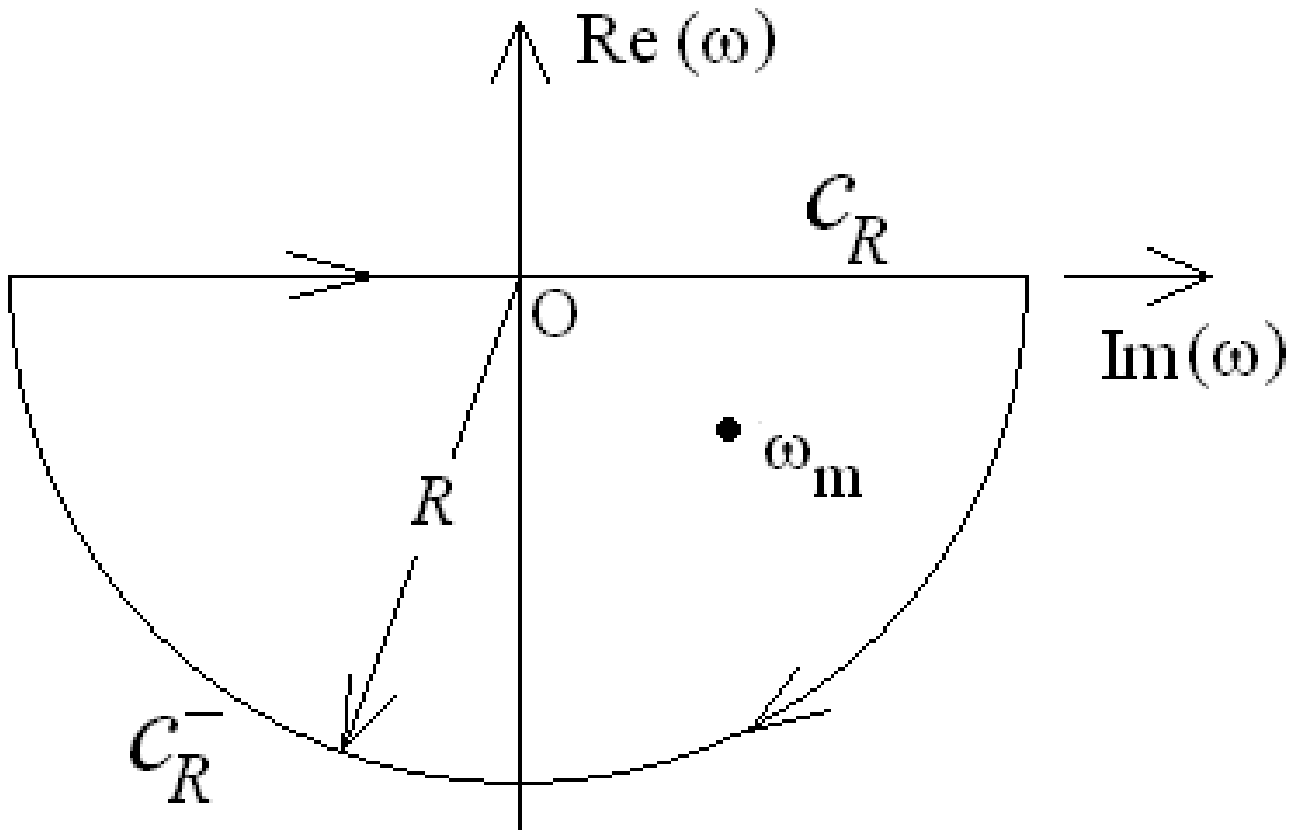}
  \caption{Contour $\mathcal{C}^{-}$.}
  \label{cminus}
  \end{center}
\end{figure}

Consequently, the time history of displacement in the layer is
\begin{equation}\label{cfpr40}
  F_{1}(t)\approx i\sum_{m\in\mathbb{Z}}\frac{1
}{\dot{D}(\omega_{m}^{o},s^{i})}
  \int_{-\infty}^{\infty}\frac{1}{\omega^{'}-\omega_{m}^{o}}
  \exp\left( -i\omega^{'}t\right)
  d\omega^{'}~.
\end{equation}
Thus, we are faced with the problem of the evaluation of the
integral
\begin{equation}\label{cfpr43}
  I(\mathbf{x},t)=
  \int_{-\infty}^{\infty}\frac{1}{\omega^{'}-\omega_{m}^{o}}
\exp\left( -i\omega t\right)
  d\omega^{'}~.
\end{equation}
wherein the important property to note is that
$\omega_{m}^{o"}<0~;~\forall m\in\mathbb{Z}$ which means that the
pole of the
  integrand lies in the lower half part of the complex $\omega$
  plane for all $m\in\mathbb{Z}$. Thus, in order to apply Cauchy's
  theorem, we consider the two contour integrals
\begin{equation}\label{cfpr44}
  I_{\mathcal{C}^{\pm}}(\mathbf{x},t)=
  \int_{\mathcal{C}^{\pm}}\frac{1}{\omega-\omega_{m}^{o}}
\exp(-i\omega t)
  d\omega~,
\end{equation}
wherein the contours $\mathcal{C}^{\pm}$ are depicted in figs.
\ref{cplus} and \ref{cminus}.
Then
\begin{equation}\label{cfpr45}
  I_{\mathcal{C}^{+}}(\mathbf{x},t)=
  \int_{\mathcal{C}_{R}^{+}}\frac{1}{\omega-\omega_{m}^{o}}
\exp(-i\omega t) d\omega +
\int_{\mathcal{C}_{R}}\frac{1}{\omega-\omega_{m}^{o}}
\exp(-i\omega t)
  d\omega~,
\end{equation}
\begin{equation}\label{cfpr45a}
  I_{\mathcal{C}^{-}}(\mathbf{x},t)=
  \int_{\mathcal{C}_{R}^{-}}\frac{1}{\omega-\omega_{m}^{o}}
\exp(-i\omega t) d\omega+
\int_{\mathcal{C}_{R}}\frac{1}{\omega-\omega_{m}^{o}}
\exp(-i\omega t)
  d\omega~.
\end{equation}
However
\begin{equation}\label{cfpr45b}
 I(\mathbf{x},t)=\lim_{R\rightarrow\infty}\int_{\mathcal{C}_{R}}\frac{1}{\omega-\omega_{m}^{o}}
\exp(-i\omega t)
  d\omega
~,
\end{equation}
so that, by virtue of Cauchy's theorem
\begin{equation}\label{cfpr46}
 \lim_{R\rightarrow\infty} I_{\mathcal{C}^{+}}(\mathbf{x},t)=\lim_{R\rightarrow\infty}
  \int_{\mathcal{C}_{R}^{+}}\frac{1}{\omega-\omega_{m}^{o}}
\exp(-i\omega t) d\omega +I(\mathbf{x},t)=0 ~,
\end{equation}
\begin{equation}\label{cfpr47}
 \lim_{R\rightarrow\infty} I_{\mathcal{C}^{-}}(\mathbf{x},t)=\lim_{R\rightarrow\infty}
  \int_{\mathcal{C}_{R}^{-}}\frac{1}{\omega-\omega_{m}^{o}}
\exp(-i\omega t) d\omega+I(\mathbf{x},t)=-2\pi
i~\text{Residue}\Big|_{\omega=\omega_{m}^{o}} ~,
\end{equation}
However:
\begin{multline}\label{cfpr48}
  \Big\|\int_{\mathcal{C}_{R}^{+}}\frac{1}{\omega-\omega_{m}^{o}}
\exp(-i\omega t) d\omega \Big\|<\int_{0}^{\pi} R\exp\left[ Rt
\sin\theta\right]d\theta\Big\| ~~;
\\
~~t<0~~,~~R>>1
 ~,
\end{multline}
so that (due to the fact that $\sin\theta\geq 0$ for
$\theta\in[0,\pi]$)
\begin{equation}\label{cfpr49}
\lim_{R\rightarrow\infty}
  \int_{\mathcal{C}_{R}^{+}}\frac{1}{\omega-\omega_{m}^{o}}
\exp(-i\omega t) d\omega=0~~;~~t<0~,
\end{equation}
whence (by virtue of (\ref{cfpr46})
\begin{equation}\label{cfpr50}
I(\mathbf{x},t)=0~~;~~t<0 ~,
\end{equation}
Similarly
\begin{multline}\label{cfpr51}
  \Big\|\int_{\mathcal{C}_{R}^{-}}\frac{1}{\omega-\omega_{m}^{o}}
\exp(-i\omega t)d\omega \Big\|<\int_{2\pi}^{\pi} R\exp\left[ Rt
\sin\theta\right]d\theta\Big\| ~~;
\\
~~t>0~~,~~R>>1
 ~,
\end{multline}
so that (due to the fact that $\sin\theta\leq 0$ for
$\theta\in[\pi,2\pi]$)
\begin{equation}\label{cfpr52}
\lim_{R\rightarrow\infty}
  \int_{\mathcal{C}_{R}^{-}}\frac{1}{\omega-\omega_{m}^{o}}
\exp\left( -i\omega t\right) d\omega=0~~;~~t>0~,
\end{equation}
whence (by virtue of (\ref{cfpr47})
\begin{equation}\label{cfpr53}
I(\mathbf{x},t)=-2\pi
i~\text{Residue}\Big|_{\omega=\omega_{m}^{o}}=-2\pi i
\exp(-i\omega t)~~;~~t>0 ~.
\end{equation}
Thus,
\begin{equation}\label{cfpr54}
I(\mathbf{x},t)=-2\pi i \exp(-i\omega_{m}^{o} t) H(t) ~
\end{equation}
wherein $H$ is the Heaviside function ($H(\chi)=0~;~\chi<0$ and
$H(\chi)=1~;~\chi>0$).

It follows from (\ref{cfprc14}) that
\begin{equation}\label{cfprc17}
 F_{1}(t_{1})=2\pi\sum_{m\in\mathbb{Z}}\frac{\exp(-i\omega_{m}^{o}t_{1})}{\dot{D}(\omega_{m}^{o},s^{i})}
 H(t_{1})
  ~.
\end{equation}
Due to the fact that $H(-t_{1})=0~;~t_{1}>0$,
\begin{equation}\label{cfprc18}
 F_{1}(-t_{1})=0~~;~~t_{1}>0~,
\end{equation}
whence
\begin{equation}\label{cfprc19}
  u_{3}^{1\pm}(\mathbf{x},t) = \frac{1}{2\pi}\int_{0}^{\infty}F_{1}(t_{1})
  F_{2}\left( \tau^{1\pm}(\mathbf{x},t,s^{i})-t_{1}\right)dt_{1}
  ~,
\end{equation}
or, more explicitly
\begin{multline}\label{cfprc20}
  u_{3}^{1\pm}(\mathbf{x},t) =\mathcal{A}\sum_{m\in\mathbb{Z}}\frac{1}{\dot{D}(\omega_{m}^{o},s^{i})}\int_{0}^{\infty}
   \left[
   -1+2\alpha^{2}(\tau^{1\pm}(\mathbf{x},t,s^{i})-t_{1}-\beta)^{2}\right]\times
   \\
\exp[-i\omega_{m}^{o}t_{1}-\alpha^{2}(\tau^{1\pm}(\mathbf{x},t,s^{i})-t_{1}-\beta)^{2}]
dt_{1}
  ~.
\end{multline}
At this point, we recall that (with obvious shorthand notation):
\begin{equation}\label{cfprc21}
 \omega_{m}^{o}= \omega_{m}^{o'}+i
 \omega_{m}^{o"}~~,~~\omega_{m}^{o'}=(2m+1)\frac{\pi}{2\gamma}~~,~~
 \omega_{m}^{o"}=\frac{1}{2\gamma}\ln\left(
 \frac{a-b}{a+b}\right)= \omega^{o"}<0
  ~,
\end{equation}
\begin{equation}\label{cfprc22}
 \dot{D}(\omega_{m}^{o},s^{i})=i(-1)^{m}\gamma\frac{b^{2}-a^{2}}{a}\cosh(\gamma\omega_{m}^{o"})=
i(-1)^{m}\mathrm{\dot{D}}(s^{i})~~,~~\mathrm{\dot{D}}(s^{i})=\gamma\frac{b^{2}-a^{2}}{a}\cosh(\gamma\omega^{o"})
  ~.
\end{equation}
\newline
{\it Remark}
\newline
We notice that $\omega_{m}^{o"}=\omega^{o"}$ and
$\mathrm{\dot{D}}(s^{i})$ are independent of $m$.
\newline
\newline
Thus, we can write (\ref{cfprc20}) as
\begin{multline}\label{cfprc23}
  u_{3}^{1\pm}(\mathbf{x},t)
  =\frac{\mathcal{A}}{i\mathrm{\dot{D}}}\int_{0}^{\infty}S(t_{1})
   \left[
   -1+2\alpha^{2}(\tau^{1\pm}(\mathbf{x},t,s^{i})-t_{1}-\beta)^{2}\right]\times
   \\
\exp[-i\omega_{m}^{o}t_{1}-\alpha^{2}(\tau^{1\pm}(\mathbf{x},t,s^{i})-t_{1}-\beta)^{2}]
dt_{1}
  ~,
\end{multline}
wherein
\begin{equation}\label{cfprc24}
 S(t_{1})=\sum_{m=-\infty}^{\infty}(-1)^{m}\exp(-i\omega_{m}^{o}t_{1})=
 \exp(\omega^{o"}t_{1})\sum_{m=-\infty}^{\infty}(-1)^{m}\exp\left[
 -i(2m+1)\frac{\pi}{2\gamma}t_{1}\right]
  ~.
\end{equation}
Thus:
\begin{multline}\label{cfprc25}
  u_{3}^{1\pm}(\mathbf{x},t)
  =\frac{\mathcal{A}}{i\mathrm{\dot{D}}}\int_{0}^{\infty}\sigma(t_{1})
   \left[
   -1+2\alpha^{2}(\tau^{1\pm}(\mathbf{x},t,s^{i})-t_{1}-\beta)^{2}\right]\times
   \\
\exp[\omega^{o"}t_{1}-\alpha^{2}(\tau^{1\pm}(\mathbf{x},t,s^{i})-t_{1}-\beta)^{2}]
dt_{1}
  ~,
\end{multline}
wherein
\begin{equation}\label{cfprc26}
 \sigma(t_{1})=\sum_{m=-\infty}^{\infty}(-1)^{m}\exp\left[
 -i(2m+1)\frac{\pi}{2\gamma}t_{1}\right] =\exp\left(
 -i\frac{\pi}{2\gamma}t_{1}\right) \sum_{m=-\infty}^{\infty}\exp\left[
 i\frac{m\pi}{\gamma}(\gamma-t_{1})\right]
  ~.
\end{equation}
We make use of the {\it Poisson sum formula} (Morse and Feshbach,
1953)
\begin{equation}\label{cfprc27}
 \sum_{m=-\infty}^{\infty}\exp\left( imdx
\right) =\frac{2\pi}{d}\sum_{m=-\infty}^{\infty}\delta\left(
x+\frac{2m\pi}{d} \right)
 ~,
\end{equation}
wherein we take $d=\frac{\pi}{\gamma}$ and $x=\gamma-t_{1}$, to
obtain
\begin{equation}\label{cfprc28}
 \sum_{m=-\infty}^{\infty}\exp\left[
 i\frac{m\pi}{\gamma}(\gamma-t_{1})\right] =2\gamma
\sum_{m=-\infty}^{\infty}\delta\left( \gamma-t_{1}+2m\gamma\right)
=2\gamma \sum_{m=-\infty}^{\infty}\delta\left(
t_{1}-(2m+1)\gamma\right)
 ~.
\end{equation}
But (recall that $\gamma>0$) $t_{1}=(2m+1)\gamma<0$ for $m<0$, so
that
\begin{equation}\label{cfprc29}
 \sum_{m=-\infty}^{\infty}\exp\left[
 i\frac{m\pi}{\gamma}(\gamma-t_{1})\right] =2\gamma
\sum_{m=0}^{\infty}\delta\left( \gamma-t_{1}+2m\gamma\right)
~~;~~t_{1}\geq 0
 ~,
\end{equation}
whence
\begin{equation}\label{cfprc30}
 \sigma(t_{1})=2\gamma\exp\left(
 -i\frac{\pi}{2\gamma}t_{1}\right) \sum_{m=0}^{\infty}\delta\left(
t_{1}-(2m+1)\gamma\right)~~;~~t_{1}\geq 0
 ~,
\end{equation}
or, on account of the properties of the Dirac delta distribution,
\begin{multline}\label{cfprc31}
 \sigma(t_{1})=2\gamma\sum_{m=0}^{\infty}\exp\left(
 -i\frac{\pi}{2\gamma}(2m+1)\gamma\right) \delta\left(
t_{1}-(2m+1)\gamma\right) =
\\
-2i\gamma\sum_{m=0}^{\infty}(-1)^{m}\delta\left(
t_{1}-(2m+1)\gamma\right)=~~;~~t_{1}\geq 0
 ~.
\end{multline}
Consequently:
\begin{multline}\label{cfprc32}
  u_{3}^{1\pm}(\mathbf{x},t)
  =\frac{-2i\gamma \mathcal{A}}{i\mathrm{\dot{D}}}\sum_{m=0}^{\infty}(-1)^{m}\int_{0}^{\infty}
  \delta\left(
t_{1}-(2m+1)\gamma\right) \left[
   -1+2\alpha^{2}(\tau^{1\pm}(\mathbf{x},t,s^{i})-t_{1}-\beta)^{2}\right]\times
   \\
\exp[\omega^{o"}t_{1}-\alpha^{2}(\tau^{1\pm}(\mathbf{x},t,s^{i})-t_{1}-\beta)^{2}]
dt_{1}
  ~,
\end{multline}
or finally, on account of the sifting property of the Dirac delta
distribution,
\begin{multline}\label{cfprc33}
  u_{3}^{1\pm}(\mathbf{x},t)
  =\frac{-2\gamma \mathcal{A}}{\mathrm{\dot{D}}}\sum_{m=0}^{\infty}(-1)^{m}
  \left[
   -1+2\alpha^{2}\{\tau^{1\pm}(\mathbf{x},t,s^{i})-(2m+1)\gamma-\beta\}^{2}\right]\times
   \\
\exp[\omega^{o"}(2m+1)\gamma-\alpha^{2}\{\tau^{1\pm}(\mathbf{x},t,s^{i})-(2m+1)\gamma-\beta\}^{2}]
  ~.
\end{multline}
The terms of the series decrease exponentially with $m$ so that
the series can be approximated by a sum of $M+1$ terms
\begin{multline}\label{cfprc34}
  u_{3}^{1\pm}(\mathbf{x},t)
  \approx\frac{-2\gamma \mathcal{A}}{\mathrm{\dot{D}}}\sum_{m=0}^{M}(-1)^{m}
  \left[
   -1+2\alpha^{2}\{\tau^{1\pm}(\mathbf{x},t,s^{i})-(2m+1)\gamma-\beta\}^{2}\right]\times
   \\
\exp[\omega^{o"}(2m+1)\gamma-\alpha^{2}\{\tau^{1\pm}(\mathbf{x},t,s^{i})-(2m+1)\gamma-\beta\}^{2}]
  ~,
\end{multline}
which is the form adopted in the numerical applications of this
method.
\newline
\newline
{\it Remark}
\newline
As shown in the following section, the pole-residue convolution
method gives rise to the correct solution in all cases.
\subsubsection{Comparison of the three  methods for evaluating the
Fourier transform intervening in the temporal response for Ricker
pulse excitation}\label{comp3c}
\begin{figure}
[ptb]
\begin{center}
\includegraphics[scale=0.45] {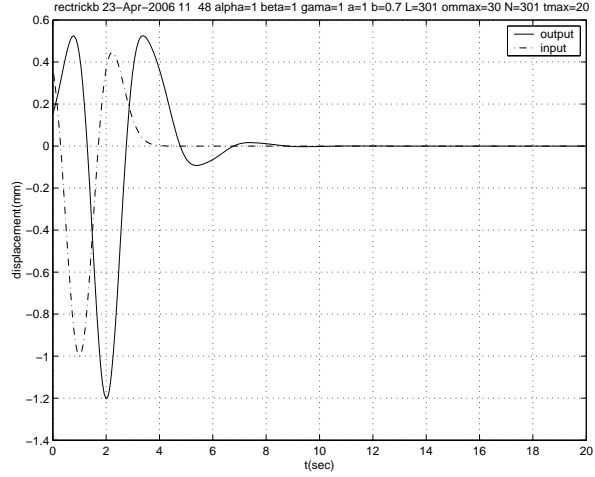}
  \caption{Time history at $\mathbf{x}=(0,0)$ computed by the rectangular quadrature method  for an "anomalous" pulse.
  $\mathcal{A}=1$, $\alpha=1$, $\beta=1$, $\gamma=1$,
  $a=1$, $b=0.7$, $\omega_{max}=30Hz$, $L=301$.}
  \label{rectrickb16}
  \end{center}
\end{figure}
\begin{figure}
[ptb]
\begin{center}
\includegraphics[scale=0.45] {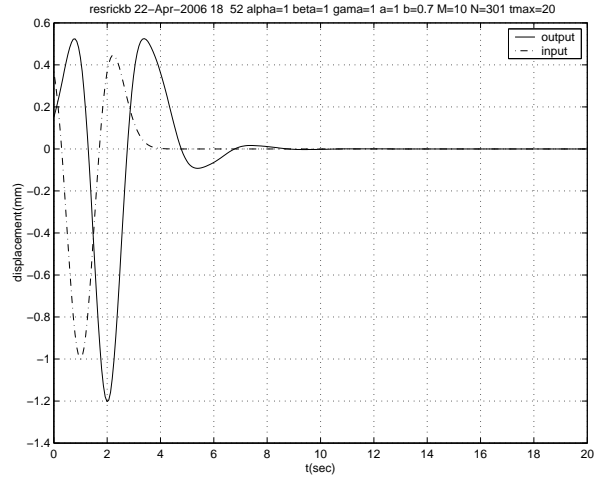}
  \caption{Time history  at $\mathbf{x}=(0,0)$ computed by the power series method  for an "anomalous" pulse.
  $\mathcal{A}=1$, $\alpha=1$, $\beta=1$, $\gamma=1$,
  $a=1$, $b=0.7$, $M=10$.}
  \label{resrickb17}
  \end{center}
\end{figure}
\begin{figure}
[ptb]
\begin{center}
\includegraphics[scale=0.45] {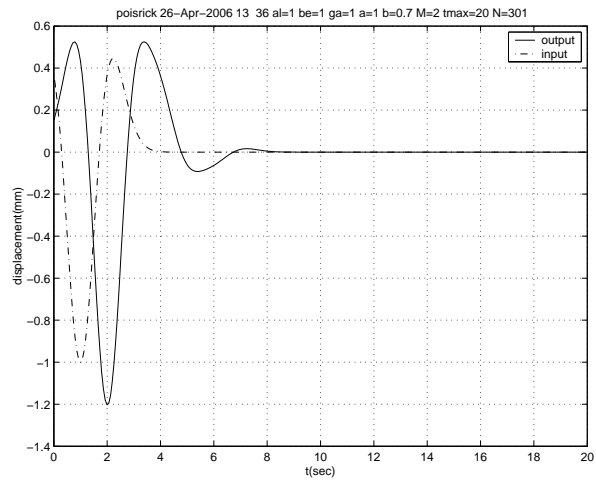}
  \caption{Time history at $\mathbf{x}=(0,0)$ computed by the pole-residue convolution method  for an "anomalous" pulse.
  $\mathcal{A}=1$, $\alpha=1$, $\beta=1$, $\gamma=1$,
  $a=1$, $b=0.7$, $M=2$.}
  \label{poisrick18}
  \end{center}
\end{figure}
\begin{figure}
[ptb]
\begin{center}
\includegraphics[scale=0.4] {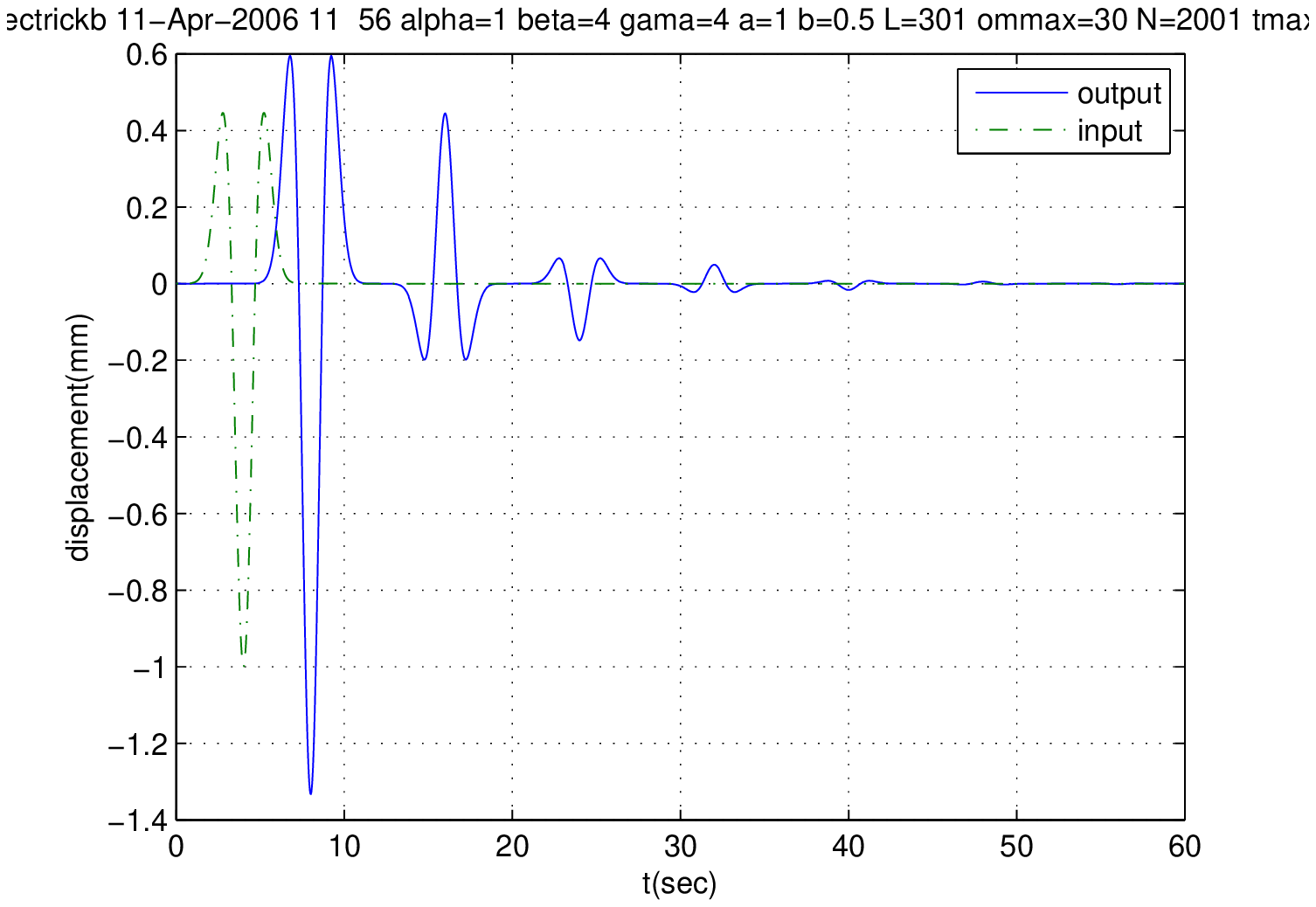}
  \caption{Time history at $\mathbf{x}=(0,0)$ computed by the rectangular quadrature method
  for separated pulses.  $\mathcal{A}=1$, $\alpha=1$, $\beta=4$, $\gamma=4$,
  $a=1$, $b=0.5$, $\omega_{max}=30Hz$, $L=301$.}
  \label{rectrickb17}
  \end{center}
\end{figure}
\begin{figure}
[ptb]
\begin{center}
\includegraphics[scale=0.4] {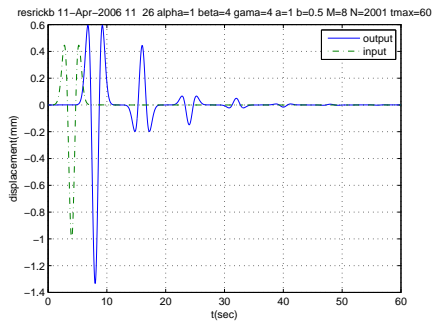}
  \caption{Time history at $\mathbf{x}=(0,0)$ computed by the power series method for separated pulses.
  $\mathcal{A}=1$, $\alpha=1$, $\beta=4$, $\gamma=4$,
  $a=1$, $b=0.5$, $M=8$.}
  \label{resrickb18}
  \end{center}
\end{figure}
\begin{figure}
[ptb]
\begin{center}
\includegraphics[scale=0.3] {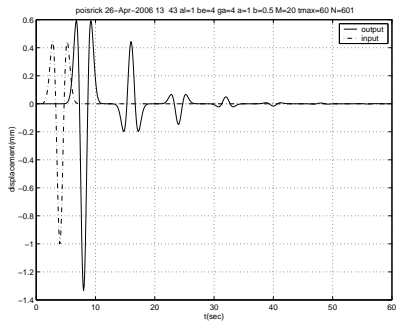}
  \caption{Time history at $\mathbf{x}=(0,0)$ computed by the pole-residue convolution method for separated pulses.
  $\mathcal{A}=1$, $\alpha=1$, $\beta=4$, $\gamma=4$,
  $a=1$, $b=0.5$,  $M=20$.}
  \label{poisrick19}
  \end{center}
\end{figure}
\begin{figure}
[ptb]
\begin{center}
\includegraphics[scale=0.3] {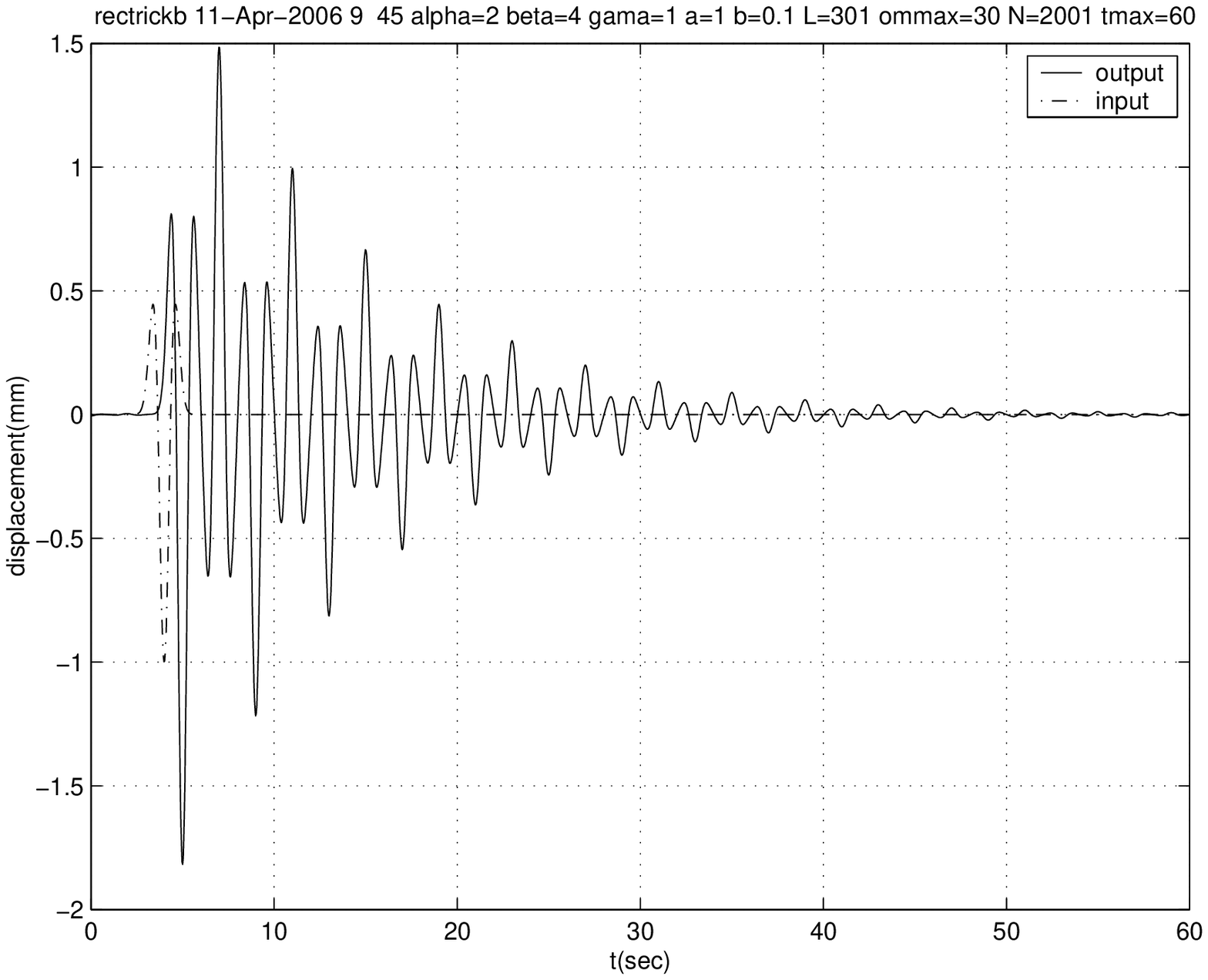}
  \caption{Time history at $\mathbf{x}=(0,0)$ computed by the rectangular quadrature method  for merged pulses.
  $\mathcal{A}=1$, $\alpha=2$, $\beta=4$, $\gamma=1$,
  $a=1$, $b=0.1$, $\omega_{max}=30Hz$, $L=301$.}
  \label{rectrickb18}
  \end{center}
\end{figure}
\begin{figure}
[ptb]
\begin{center}
\includegraphics[scale=0.3] {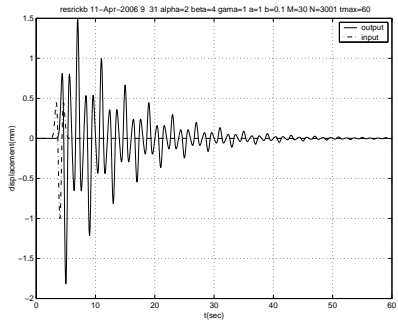}
  \caption{Time history at $\mathbf{x}=(0,0)$ computed by the power series method  for merged pulses.
  $\mathcal{A}=1$, $\alpha=2$, $\beta=4$, $\gamma=1$,
  $a=1$, $b=0.1$, $M=30$.}
  \label{resrickb19}
  \end{center}
\end{figure}
\begin{figure}
[ptb]
\begin{center}
\includegraphics[scale=0.3] {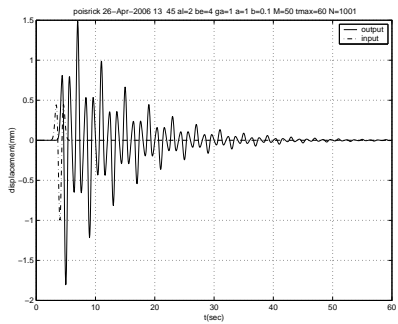}
  \caption{Time history at $\mathbf{x}=(0,0)$ computed by the pole-residue convolution method  for merged pulses.
  $\mathcal{A}=1$, $\alpha=2$, $\beta=4$, $\gamma=1$,
  $a=1$, $b=0.1$, $M=50$.}
  \label{poisrick20}
  \end{center}
\end{figure}
\begin{figure}
[ptb]
\begin{center}
\includegraphics[scale=0.9] {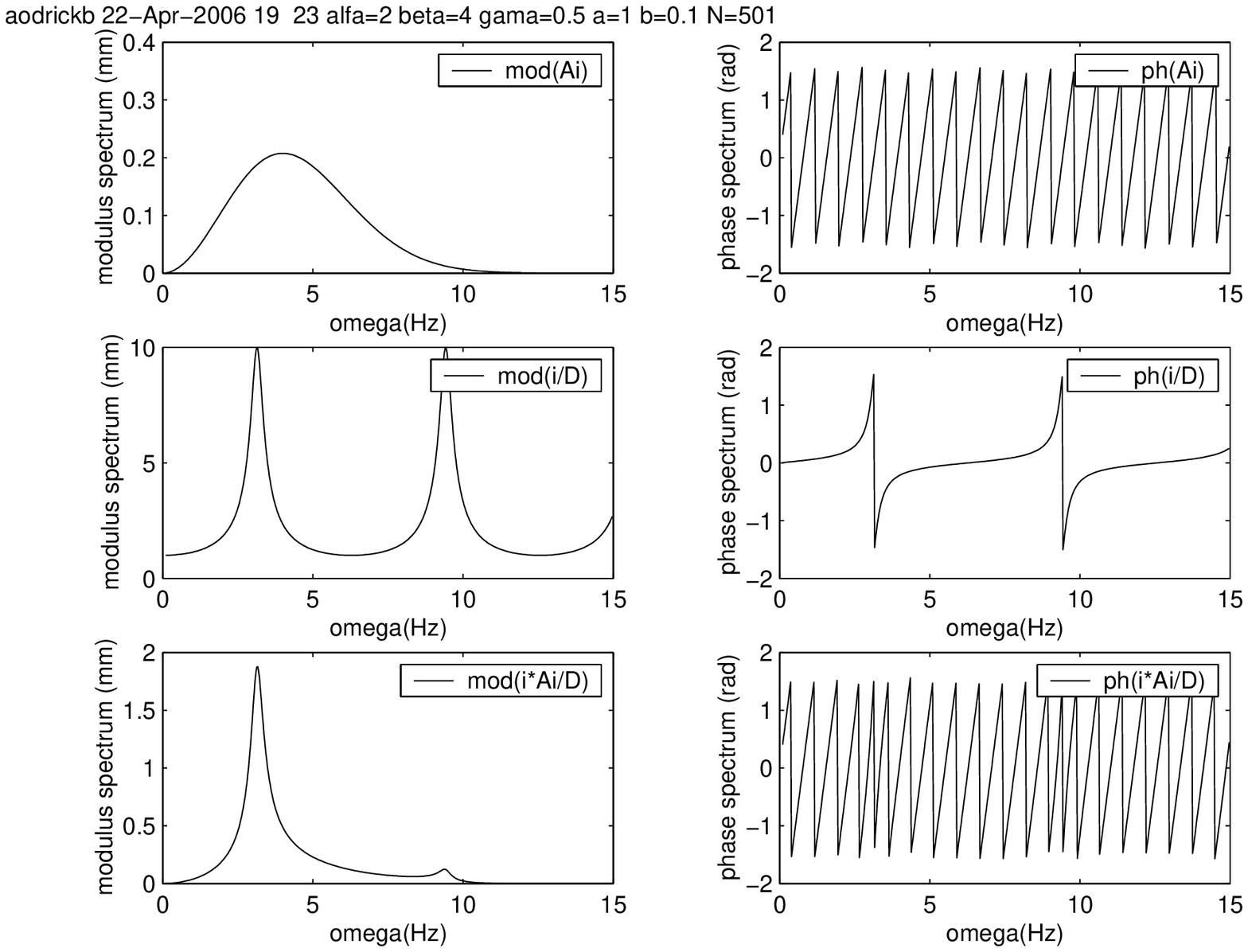}
  \caption{Spectrum of displacement response at $\mathbf{x}=(0,0)$.
  $\mathcal{A}=1$,
  $\alpha=2$, $\beta=4$, $\gamma=0.5$, $a=1$, $b=0.1$,
  corresponding to the case of a quasi-monochromatic pulse. The left-hand
  curves
  pertain to moduli, and the right hand curves to phases of the spectra.}
  \label{aodrickb10}
  \end{center}
\end{figure}
\begin{figure}
[ptb]
\begin{center}
\includegraphics[scale=0.35] {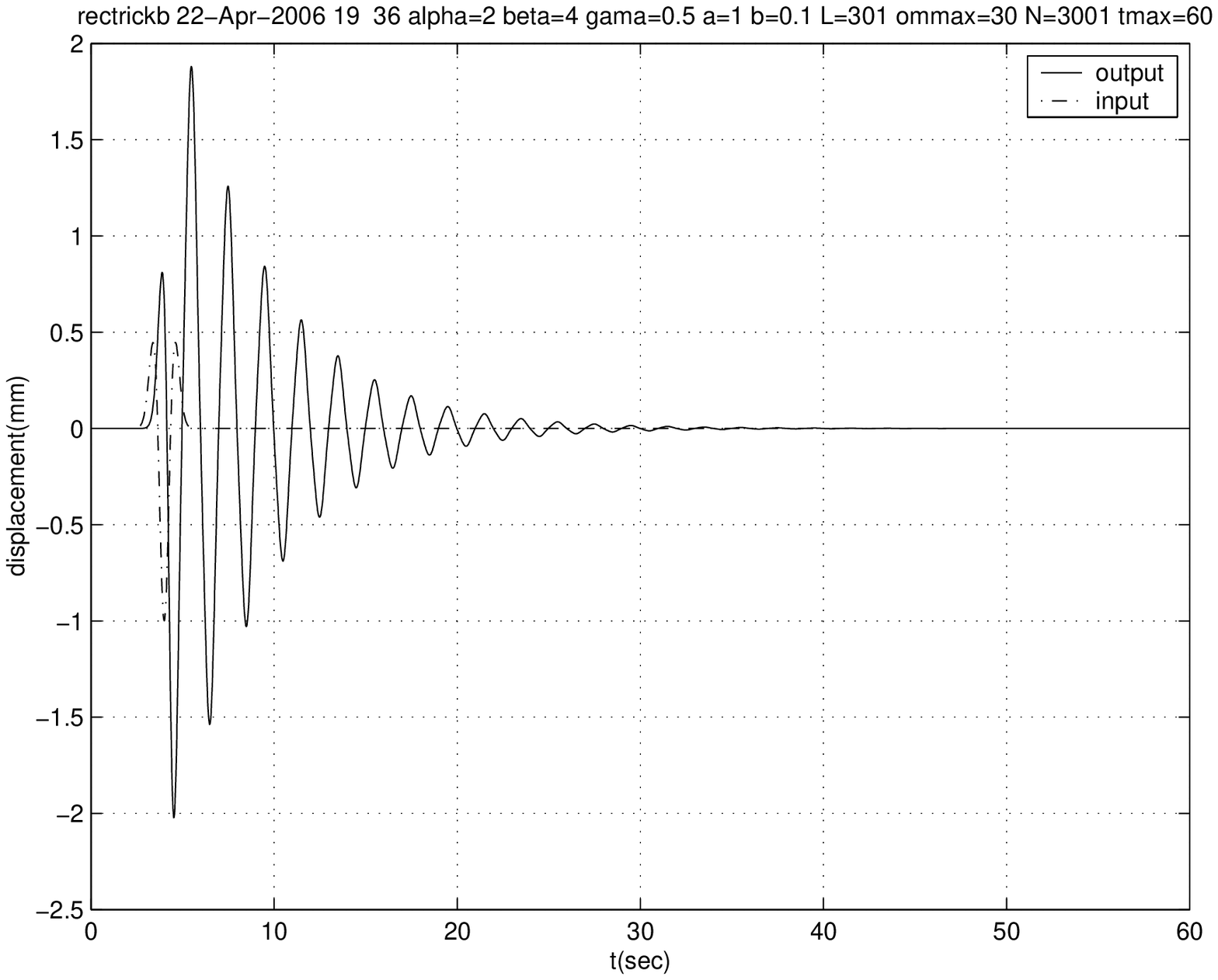}
  \caption{Time history at $\mathbf{x}=(0,0)$ computed by the rectangular quadrature method
  for  a quasi monochromatic pulse.
  $\mathcal{A}=1$, $\alpha=2$, $\beta=4$, $\gamma=0.5$,
  $a=1$, $b=0.1$, $\omega_{max}=30Hz$, $L=301$.}
  \label{rectrickb19}
  \end{center}
\end{figure}
\begin{figure}
[ptb]
\begin{center}
\includegraphics[scale=0.35] {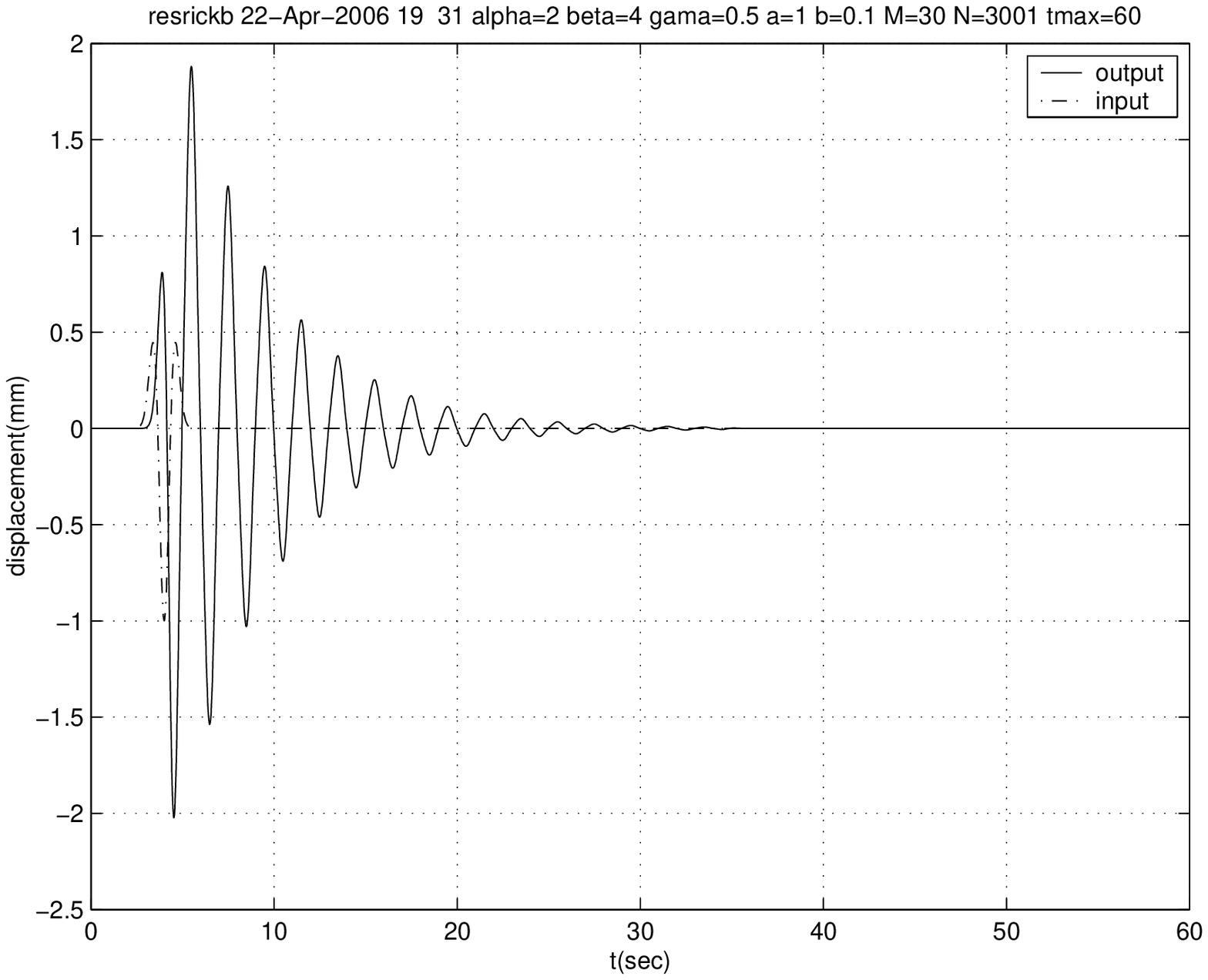}
  \caption{Time history at $\mathbf{x}=(0,0)$ computed by the power series method  for  a quasi monochromatic pulse.
  $\mathcal{A}=1$, $\alpha=2$, $\beta=4$, $\gamma=0.5$,
  $a=1$, $b=0.1$, $M=30$.}
  \label{resrickb20}
  \end{center}
\end{figure}
\begin{figure}
[ptb]
\begin{center}
\includegraphics[scale=0.475] {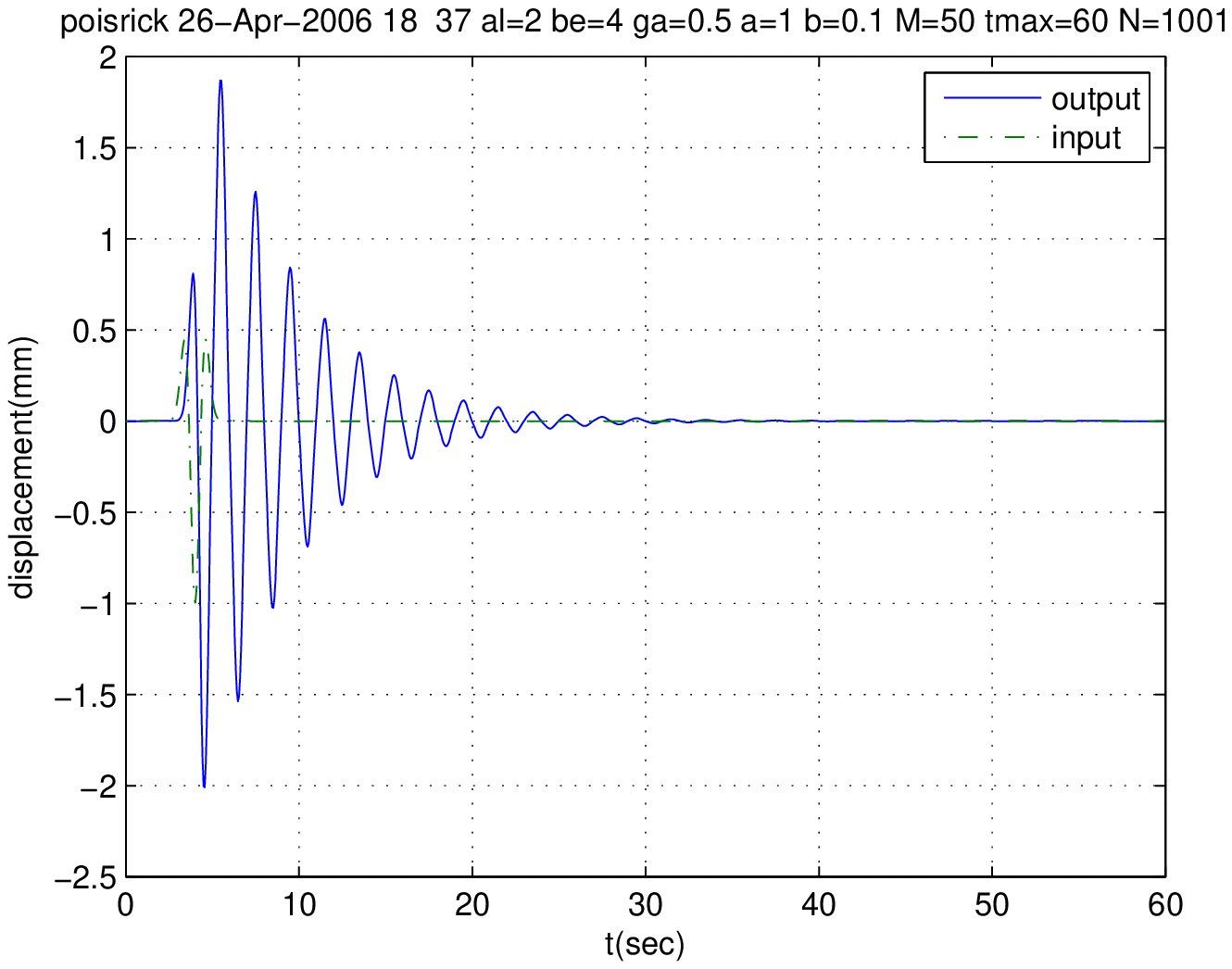}
  \caption{Time history at $\mathbf{x}=(0,0)$ computed by the pole-residue convolution method  for  a quasi monochromatic pulse.
  $\mathcal{A}=1$, $\alpha=2$, $\beta=4$, $\gamma=0.5$,
  $a=1$, $b=0.1$, $M=50$.}
  \label{poisrick21}
  \end{center}
\end{figure}
\begin{figure}
[ptb]
\begin{center}
\includegraphics[scale=0.9] {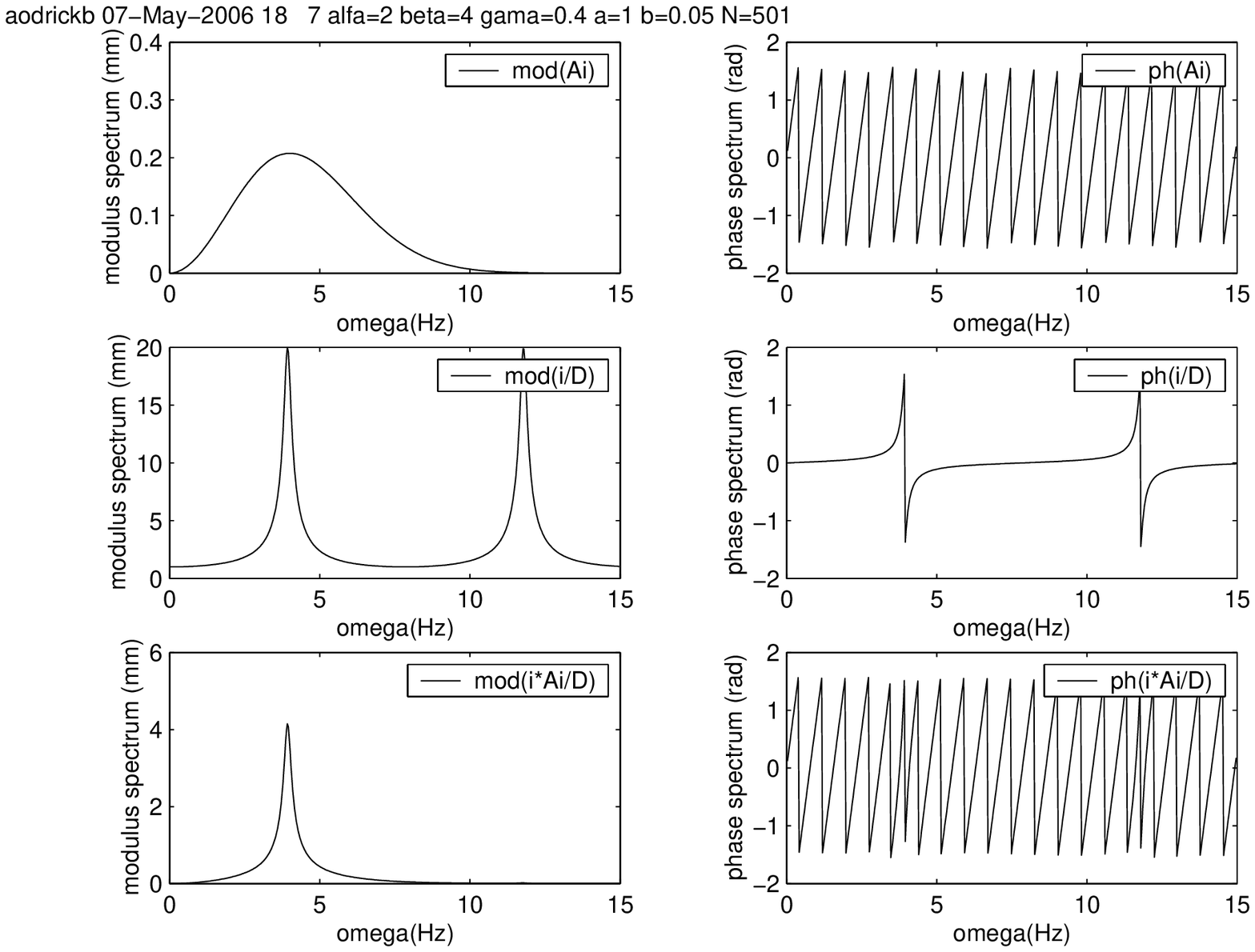}
  \caption{Spectrum of displacement response at $\mathbf{x}=(0,0)$.
  $\mathcal{A}=1$,
  $\alpha=2$, $\beta=4$, $\gamma=0.4$, $a=1$, $b=0.05$,
  corresponding to the case of a monochromatic pulse. The left-hand
  curves
  pertain to moduli, and the right hand curves to phases of the spectra.}
  \label{aodrickb11}
  \end{center}
\end{figure}
\begin{figure}
[ptb]
\begin{center}
\includegraphics[scale=0.425] {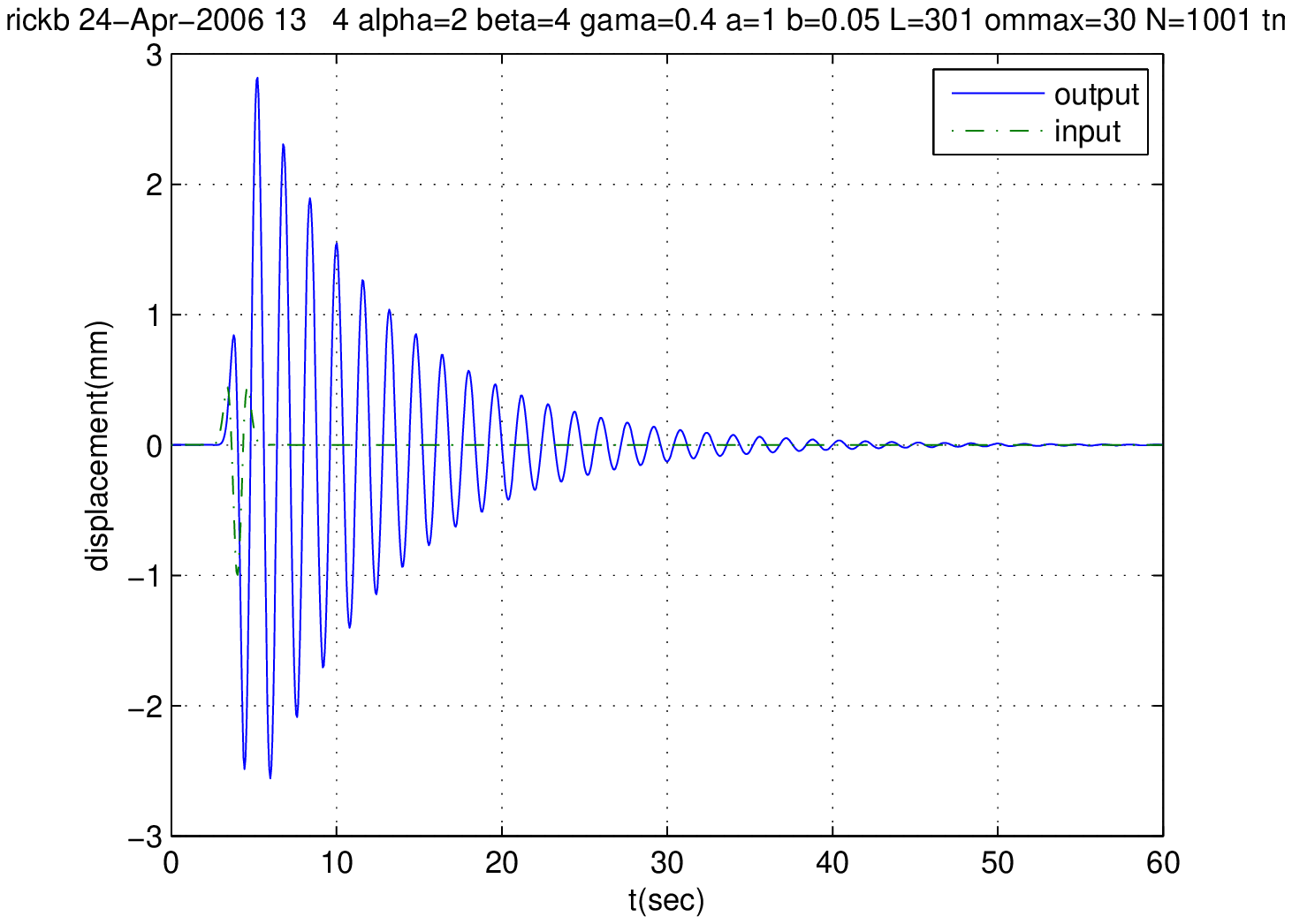}
  \caption{Time history at $\mathbf{x}=(0,0)$ computed by the rectangular quadrature method  for a  monochromatic pulse.
  $\mathcal{A}=1$, $\alpha=2$, $\beta=4$, $\gamma=0.4$,
  $a=1$, $b=0.05$, $\omega_{max}=30Hz$, $L=301$.}
  \label{rectrickb20}
  \end{center}
\end{figure}
\begin{figure}
[ptb]
\begin{center}
\includegraphics[scale=0.425] {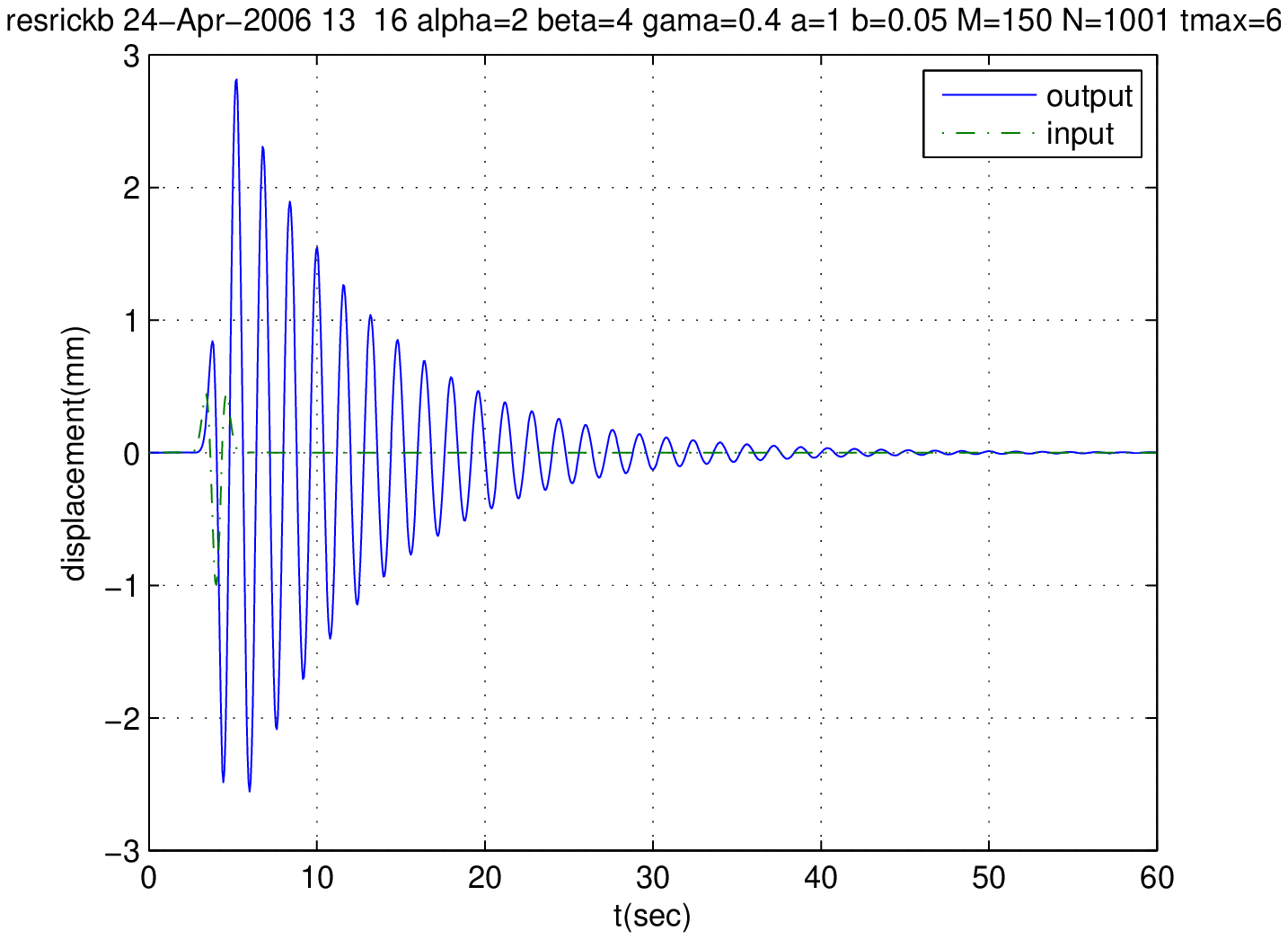}
  \caption{Time history at $\mathbf{x}=(0,0)$ computed by the power series method  for a  monochromatic pulse.
  $\mathcal{A}=1$, $\alpha=2$, $\beta=4$, $\gamma=0.4$,
  $a=1$, $b=0.05$, $M=150$.}
  \label{resrickb21}
  \end{center}
\end{figure}
\begin{figure}
[ptb]
\begin{center}
\includegraphics[scale=0.425] {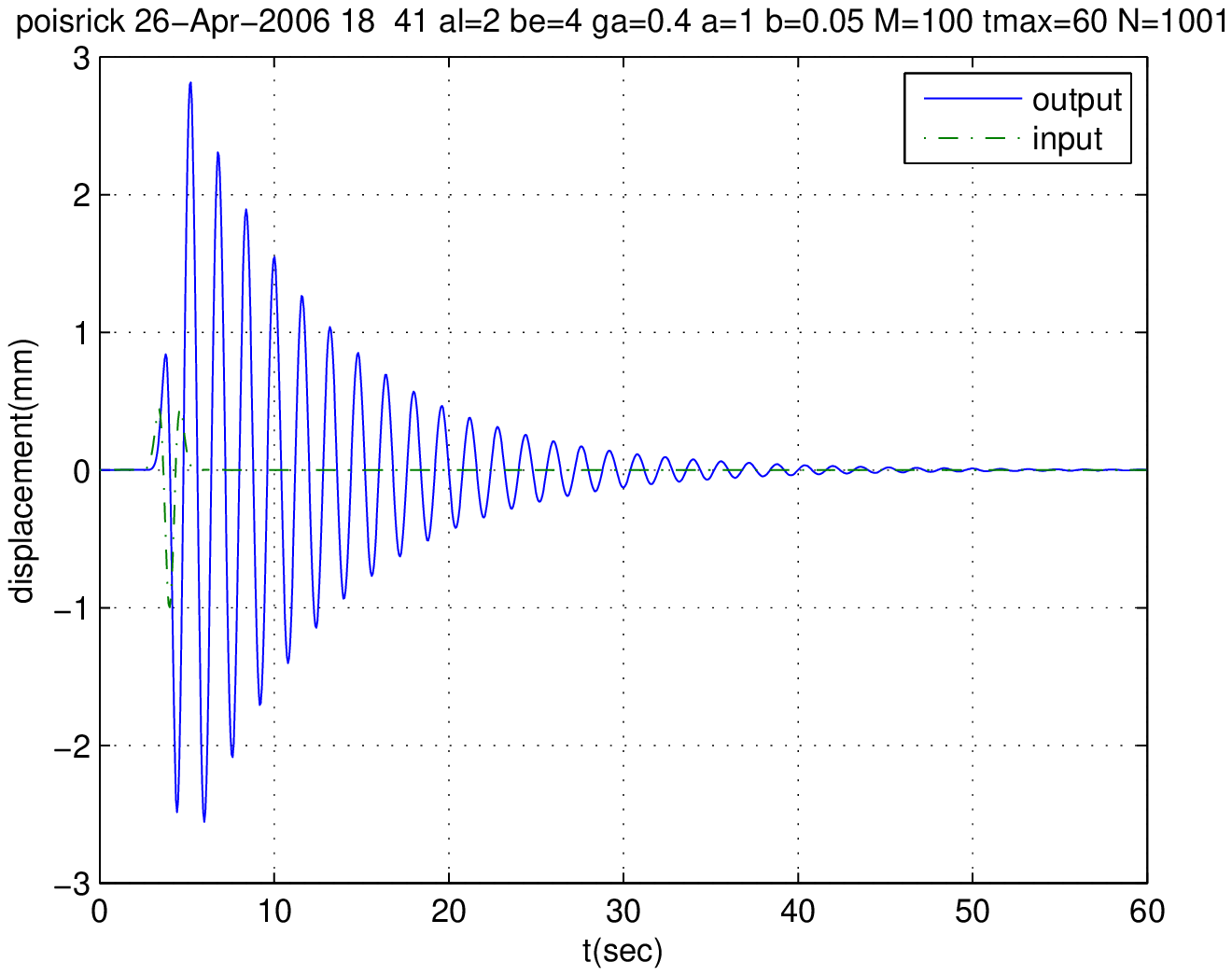}
  \caption{Time history at $\mathbf{x}=(0,0)$ computed by the pole-residue convolution method  for a  monochromatic pulse.
  $\mathcal{A}=1$, $\alpha=2$, $\beta=4$, $\gamma=0.4$,
  $a=1$, $b=0.05$, $M=100$.}
  \label{poisrick22}
  \end{center}
\end{figure}
The three methods are: i) the rectangle quadrature method, ii) the
power series method, and iii) the complex frequency pole-residue
convolution method.

In the set of figures \ref{rectrickb16}, \ref{resrickb17},
\ref{poisrick18} (for a so-called "anomalous" pulse),
\ref{rectrickb17}, \ref{resrickb18}, \ref{poisrick19} (for
separated pulses), and \ref{rectrickb18}, \ref{resrickb19},
\ref{poisrick20} (for merged pulses) we exhibit the time history
of displacement response at the location $\mathbf{x}=(0,0)$ on the
ground plane.

We next choose what seems a typical geophysically-interesting
situation (at least in the frequency domain) depicted in fig.
\ref{aodrickb10}. The corresponding time-domain responses are
given in figures \ref{rectrickb19}, \ref{resrickb20},
\ref{poisrick21}.

As a last example, we choose a more idealized (and less realizable
due to the very large contrast of physical properties it implies
between the layer and the substratum) geophysical example, the
frequency response of which is depicted in fig. \ref{aodrickb11}.
The corresponding time-domain responses are given in figures
\ref{rectrickb20}, \ref{resrickb21}, \ref{poisrick22}.
\newline
\newline
{\it Remark}
\newline
We notice that all the three methods  again give the same results.
\newline
\newline
{\it Remark}
\newline
Note that the relatively-long duration and monochromatic nature of
the temporal response in figs. \ref{rectrickb20},
\ref{resrickb21}, \ref{poisrick22} are due to the large Q
single-spike nature of the frequency response, the latter being a
result of the large contrast of physical properties and the fact
that only one $1/D$ spike is located within the significant part
of the spectrum of the Ricker pulse.
\subsubsection{Discussion}
{\it Remark}
\newline
The rectangle quadrature method, embodied in (\ref{rq11}),
produces a purely-numerical result  which gives no insight as to
the physical nature of this response. It was proposed only as a
reference solution by which the other two methods could be judged,
at least on a numerical basis. This rectangle quadrature method is
certainly not optimal, even from the purely-numerical point of
view, but obviously one of the simplest to explain and program.
\newline
\newline
{\it Remark}
\newline
By inspection of (\ref{psq22}) and comparison with (\ref{tds158}),
we see that the power series method gives rise to an expression of
the time history response to a Ricker pulse that is a sum of
displaced (and increasingly-attenuated) Ricker pulses. This is
what one would expect on an intuitive basis for a dispersionless
configuration. Thus, it would seem that the power series method is
the most appropriate one, at least in the situation in which the
successive pulses are well-separated. However, in the case in
which the successive pulses are not well-separated, intuition is
lost (especially when a long-duration quasi-monochromatic response
is produced) and the power series picture reflects this fact,
although it still gives rise to the correct numerical response.
However, the power series method cannot be applied when $D\approx
0$ as is the case in which Love modes are excited. This is the
reason why the pole-residue convolution method was proposed.
\newline
\newline
{\it Remark}
\newline
The pole-residue convolution solution in (\ref{cfprc34}) expresses
the time history of response as a weighted  sum of displaced
Ricker pulses (the latter would probably be distorted Ricker
pulses in the presence of dispersion). This is close to being
intuitive, but what is less intuitive is the fact that the
displacements are a function of the real part of the complex zeros
of  the equation $D(\omega,s^{i})=0$ and the weight functions  are
expressed in terms of the imaginary part of the complex zeros of
 the equation $D(\omega,s^{i})=0$.

The fact that the essential features (peak values and duration,
amongst others) of the time history are directly-related to the
complex eigenvalues of the structure is the essential result we
were aiming at in this contribution.

The most important parameter is the imaginary part of the
eigenvalues since it regulates the height of the succesive Ricker
pulses and therefore determines the duration of the time domain
response. This parameter is a measure of {\it radiation damping}
(which is leakage of energy into the substratum, an attenuation
mechanism that exists even in the absence of material dissipation
in the layer).

The pole-residue convolution expression of the time history
appears to be similar to the one obtained by the power series
method, but the latter method is not applicable when
$D(\omega,s^{i})=0$ for real eigenvalues in the absence of
dissipation (i.e., the situation in which it is possible to excite
Love modes (Groby and Wirgin 2005a,b)); moreover, the power series
method does not enable one to predict the duration in an obvious
way.
\newline
\newline
{\it Remark}
\newline
Some of the numerical results included in this work are rather
unexpected. For instance, the time histories given in figs.
\ref{rectrickb18}, \ref{resrickb19}, \ref{poisrick20} have quite
long durations that one would not expect to occur for a case in
which modes cannot be excited. Actually, this long duration is due
to the fact that the only attenuative action in this work is the
one due to radiation damping. The duration would be shorter if
material dissipation (i.e., viscoelasticity) were taken into
account in the layer and/or the contrast between $a$ and $b$ were
smaller.
\section{Conclusions}
The main result of this contribution is that the three methods
give rise to the same solutions for a large variety of scattering
configurations.

The complex frequency pole-residue convolution method turns out to
be the most interesting method since: i) it is
numerically-efficient, ii) it is explicit as concerns the
understanding and quantification of the duration of the time
domain response, iii) it can be employed even in the case in which
genuine resonances (due to mode excitation) are produced.

 The part
of this study concerning the complex frequency pole residue
convolution method constitutes a correction of its counterpart in
our previous publications (Groby and Wirgin, 2005a) and (Groby and
Wirgin 2005b). A somewhat similar approach, although applied to a
fluid layer in a fluid host, is that of (Conoir, 1987).

Work remains to be done to take into account dispersion and
damping of the material in the layer.

The natural follow-up of this study is to elucidate theoretically
the nature of the time histories of response not only for the case
(the one treated herein) in which the configuration is unable to
excite (e.g., Love) modes, but also in the case in which such
modes can be excited (Groby and Wirgin 2005a,b).
\newpage
\section*{References}\label{refs}
%
\begin{itemize}
\item Conoir J.-M., R\'eflexion et transmission par une plaque
fluide, in {\it La Diffusion Acoustique}, Gespa N. (ed.), Cedocar
Paris Arm\'ees, Paris, 1987, 105-132.

\item Groby J.-P. and Wirgin A., 2D ground motion at a soft
viscoelastic layer/hard substratum site in response to  SH
cylindrical seismic waves radiated by near and distant line
sources. I. Theory, {\it Geophys.J.Int.}, 2005, 163, 165-191.

\item Groby J.-P. and Wirgin A., 2D ground motion at a soft
viscoelastic layer/hard substratum site in response to SH
cylindrical seismic waves radiated by near and distant line
sources. II. Computations, {\it Geophys.J.Int.}, 2005, 163,
192-224.

\item Hodgman C.D.(ed.), {\it CRC Standard Mathematical Tables},
Chemical Rubber Publ. Co., Cleveland, 1957, 304.

\item Morse P.M. and Feshbach H., {\it Methods of Theoretical
Physics}, Mc Graw-Hill, New York, 1953.

\item Sanchez-Sesma F.J, Diffraction of elastic SH waves by
wedges, {\it Bull.Seism.Soc.Am.},  1985, 75, 1435-1446.
\end{itemize}
\end{document}